%% file: main.tex
\documentclass{article}

\input{preambles/preamble}
\input{preambles/symbols}

\icmltitlerunning{Renaissance of Literate Programming in the Era of LLMs}

\begin{document}

    \input{sections/title}

    \input{sections/0_abs}
    \input{sections/1_intro}

    \input{sections/2_related_works}

    \input{sections/3_preliminary}

    \input{sections/4_our_work}

    \input{sections/5_exp}
    \input{sections/6_conclusion}

    \bibliographystyle{icml2025}
    \bibliography{references/yansong_added, references/yw_added, references/manual_added, references/duolei_added, references/noctis_added, references/yingyao_added, references/chichun_added, references/sirou_added, references/dashen_added, references/icml_goldfish}

    \newpage
    \appendix
    \onecolumn
    \input{sections/appendix}

\end{document}

%% file: preambles/preamble.tex
\usepackage[preprint]{icml2025_preprint}

\usepackage{titlesec}

\usepackage{lmodern}

\usepackage{multirow}
\usepackage{hyperref}
\usepackage[frozencache,cachedir=.]{minted}

\usepackage{xcolor}
\definecolor{darkgreen}{rgb}{0,0.5,0}
\definecolor{darkred}{rgb}{0.5,0,0}
\definecolor{goldyellow}{rgb}{0.5,0.5,0}

\usepackage[T1]{fontenc}
\usepackage[utf8]{inputenc}
\usepackage[english]{babel}
\usepackage{graphicx}
\usepackage[font=scriptsize]{subcaption} 

\usepackage{microtype}
\usepackage{booktabs} 
\usepackage{listings}

\usepackage{amsmath,amsfonts,amssymb,amsthm,bm,cleveref}
\theoremstyle{plain}

\theoremstyle{definition}

\theoremstyle{remark}

\usepackage{mathtools}
\expandafter\def\csname ver@etex.sty\endcsname{3000/12/31}

\usepackage{autonum}

\usepackage{algorithm}
\usepackage{algorithmic}
\renewcommand{\algorithmicrequire}{\textbf{Input:}}
\renewcommand{\algorithmicensure}{\textbf{Output:}}

%% file: preambles/symbols.tex
\usepackage{acro}

\DeclareAcronym{cot}{
  short = CoT,
  long = Chain of Thought,
}
\DeclareAcronym{llms}{
  short = LLMs,
  long = Large Language Models,
}
\DeclareAcronym{llm}{
  short = LLM,
  long = Large Language Model,
}

\DeclareAcronym{lp}{
  short = LP,
  long = Literate Programming,
}

\DeclareAcronym{ilp}{
  short = ILP,
  long = Interoperable Literate Programming,
}

\usepackage{newtxtext}
\newcommand{\TeXmacs}{T\kern-.1667em\lower.5ex\hbox{E}\kern-.125emX\kern-.1em\lower.5ex\hbox{\textsc{m\kern-.05ema\kern-.125emc\kern-.05ems}}}

\makeatletter
\def\LyX{L\kern-.2em
{\sbox\z@ X%
\vbox to .9\ht\z@{\hbox{\check@mathfonts
    \fontsize\sf@size\z@
    \math@fontsfalse\selectfont
    Y}%
  \vss}%
}%
\kern-.125emX\@}
\makeatother

%% file: sections/title.tex
\twocolumn[
\icmltitle{Renaissance of Literate Programming in the Era of LLMs: Enhancing LLM-Based Code Generation in Large-Scale Projects}

\icmlsetsymbol{equal}{*}

\begin{icmlauthorlist}
    \icmlauthor{Wuyang Zhang}{equal,liii}
    \icmlauthor{Yansong Li}{equal,liii,uic}
    \icmlauthor{Zeyu Dong}{sbu}
    \icmlauthor{Yu Wu}{rgs}
    \icmlauthor{Yingyao Zhou}{uic}
    \icmlauthor{Duolei Wang}{cuhksz}
    \icmlauthor{Songsirou Xing}{jhu}
    \icmlauthor{Chichun Zhou}{dali}
    \icmlauthor{Da Shen}{liii}
\end{icmlauthorlist}

\icmlaffiliation{uic}{University of Illinois Chicago}
\icmlaffiliation{liii}{Liii Network}
\icmlaffiliation{sbu}{Stony Brook University}
\icmlaffiliation{rgs}{Rutgers University}
\icmlaffiliation{jhu}{John Hopkins University}
\icmlaffiliation{cuhksz}{The Chinese University of Hong Kong, Shenzhen}
\icmlaffiliation{dali}{Dali University}

\icmlcorrespondingauthor{Yansong Li \& Da Shen}{da@liii.pro}
\icmlkeywords{Prompt Engineering}

\vskip 0.3in
]

\printAffiliationsAndNotice{\icmlEqualContribution} 

%% file: sections/0_abs.tex
\begin{abstract}


    \ac{llms} have helped programmers increase efficiency through code generation, comprehension, and repair. However, their application to large-scale projects remains challenging due to complex interdependencies and the extensive size of modern codebases. Although Knuth’s concept of \ac{lp} combines code and natural language to convey logic and intent, its potential for enhancing relationships in large projects has not been fully explored. 
    In this study, we introduce the idea of Interoperable \ac{lp} (ILP), which leverages literate programming principles to enhance the development of both small-scale documents and large-scale projects with \ac{llms}. We investigate how \ac{llms} perform under ILP-style instructions for both document-oriented tasks and entire projects. Recognizing that many researchers rely on well-structured templates to guide \ac{llms}, we propose a concise prompt-engineering way to write \ac{lp} documents so \ac{llms} can better be involved in code generation. We also examine the capacity of various \ac{llms} to generate Scheme and Python code on the RepoBench benchmark, illustrating the advantages of our approach. Our findings indicate that ILP with \ac{llms} can enhance LLM-based code generation in large-scale project development.

\end{abstract}

%% file: sections/1_intro.tex


\section{Introduction} \label{sec:intro}

Generative machine learning models, particularly large language models (LLMs), have significantly enhanced how programmers approach coding tasks. These models excel at generating functional code snippets, adding detailed comments, and performing tasks like code translation, repair, and test generation~\citep{jiang2024survey, chen2021evaluating}. For example, LLMs can translate user prompts into working implementations while simultaneously offering clear explanations of the code’s logic, making programming more accessible to users with varying levels of expertise. This dual capability of generating and documenting code has positioned LLMs as indispensable tools for modern software development.

Currently, LLMs are widely used for two main purposes: efficient code generation~\citep{jiang2024survey} and comprehensive code explanation~\citep{nam2024using}. Traditional code generation focuses on creating single pieces of code based on user-provided prompts. These prompts can range from detailed, step-by-step instructions guiding the implementation process to high-level descriptions with sample inputs and outputs that define the behavior of an API. While effective for generating small and isolated pieces of code, this method is inherently limited in scope, as it treats each task as independent and does not account for broader project contexts. However, as software projects grow in scale and complexity, traditional approaches to code generation become increasingly inadequate~\citep{dou2024s}.

Designing APIs or generating multiple interdependent components in large-scale projects demands more than isolated, prompt-based workflows. Such projects require a framework that manages complex interdependencies between APIs, ensures consistency across components, and supports interconnected development processes. To address these challenges, a new framework is necessary—one that facilitates bulk API generation handles dependencies, and enables the creation of components reliant on yet-to-be-developed APIs. This approach would provide the structure and coherence needed to tackle the intricacies of modern, large-scale software projects.


To improve LLM performance in generating and explaining code for large projects, we revisit the concept of \ac{lp}. \ac{lp}, while established, still offers a foundation for organizing interdependencies and enforcing consistent standards across components. Tools like Jupyter Notebook and NoWEB demonstrate the benefits of combining code and documentation within a single environment (Fig.~\ref{fig:compare_lp_with_jupyter}). However, they lack universal standards and do not fully address the demands of modern software development. In particular, they do not offer robust methodologies for generating and explaining code across interdependent systems.


In this paper, we build upon \ac{lp} to propose a new standard called \ac{ilp}. \ac{ilp} provides a structured approach tailored to LLM-based code generation and explanation. By defining explicit guidelines for code logic, \ac{ilp} enhances accuracy and development efficiency. We also introduce two software tools—Mogan and Goldfish Scheme—that showcase practical applications of \ac{ilp}. These tools help developers manage large projects, integrate documentation, and test code within a unified and secure environment.

This paper addresses the limitations of traditional code generation methods by introducing a framework based on \ac{ilp}. Experiments are conducted in RepoBench~\citep{liu_repobench_2023} on Python code generation tasks. Our key contributions are:
\begin{itemize}
    \item Our evaluations show that \ac{ilp} reduces variability in code generation quality. The performance remains consistent across instruct-based models, including locally deployed pre-trained models and online models with internet search capabilities.
    \item \ac{ilp} improves code generation efficiency, achieving speed increases that lead to higher productivity.
    \item \ac{ilp} supports bulk API generation and resolves complex interdependencies, including those involving components yet to be developed. This approach addresses critical challenges in scaling code generation for large systems.
    \item We provide comprehensive benchmarks and evaluations, rather than a single metric of execution success, to demonstrate \ac{ilp}’s effectiveness across different coding environments. These benchmarks compare code quality, speed, and completion rates across multiple models.
\end{itemize}

%% file: sections/2_related_works.tex
\paragraph{Related works}

\emph{Literate programming supported editor}:
In this paper, we introduce \href{https://mogan.app/}{Mogan}, a UTF-8-compatible literate programming editor that allows users to edit and export all project files within a single document, as shown in Figures~\ref{fig:compare_lp_with_jupyter} and \ref{fig:mogan_project_export}. This unified approach aims to simplify the workflow for large-scale projects. Other literate programming editors are discussed in Section~\ref{subsec:lp-editor}.

\emph{Prompt engineering}:
\citet{wei_chain_2022} propose the chain-of-thought (\ac{cot}) method, guiding large language models (LLMs) to reason step by step. This improves both performance and interpretability. Building on this idea, least-to-most (LtM) prompting~\citep{zhou2022least} divides complex tasks into sequential subproblems, helping LLMs tackle more difficult scenarios. Recent work also focuses on increasing consistency and accuracy. For example, \citet{wang2022self} introduce self-consistency, a strategy that produces multiple reasoning paths from a single prompt and selects the most coherent outcome. \citet{dhuliawala2023chain} present Chain-of-Verification (CoVE), a method that prompts LLMs to verify their outputs, reducing hallucinations. To reduce the manual effort in crafting prompts, researchers have explored automated prompt construction~\citep{zhou2022large, pryzant2023automatic, ye2023prompt}. These methods guide the LLM to examine its responses and propose improvements, as shown by \citet{li2023chain}, who refine the ``think step by step'' directive by prompting the LLM to identify its own limitations and update its solution accordingly.

\emph{Scheme language}:
\citet{sussman_scheme_1975} implement ``Scheme'', a simple and lightweight functional programming language based on lambda calculus. Furthermore, several Scheme interpreters have been developed based on the Scheme standard, with the latest and most widely accepted being \href{https://r7rs.org/}{R7RS}. Additionally, Scheme Requests for Implementation (\href{https://srfi.schemers.org/}{SRFI}) are community-driven specifications that extend the language with additional features and libraries. The \href{https://github.com/LiiiLabs/goldfish}{Goldfish Scheme~\citep{goldfish}}, 
%
%
built on S7 Scheme, supports the R7RS standard, parts of SRFI, and many functionalities inspired by the Python standard library. This simplicity, combined with several characteristics of Python, makes it easier for large language models to understand.
More discussion about Scheme interpreters is in Sec.~\ref{app:scheme-language}.

%% file: sections/3_preliminary.tex
\section{Background} \label{sec:prelim}




In this section, we provide background on \ac{llm}-based code generation, \ac{lp}, and the Goldfish Scheme, a Scheme interpreter developed via the \ac{lp} paradigm.

\begin{figure*}[ht]
   \centering
   \begin{subfigure}[b]{0.5\textwidth}
      \centering
      \includegraphics[width=\textwidth]{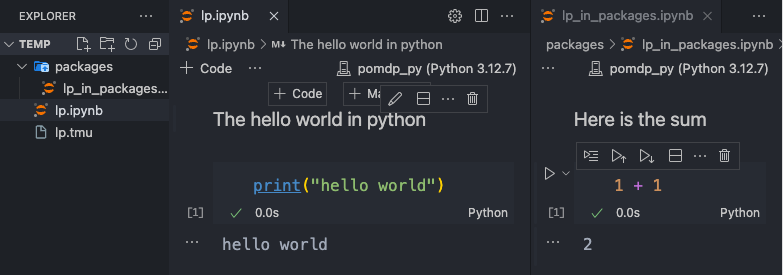}
      \caption{Two files \texttt{packages/lp\_in\_packages.ipynb} and \texttt{lp.ipynb} are edited in different documents}
      \label{subfig:jupyter}
   \end{subfigure}
   \hfill
   \begin{subfigure}[b]{0.4\textwidth}
      \centering
      \includegraphics[width=\textwidth]{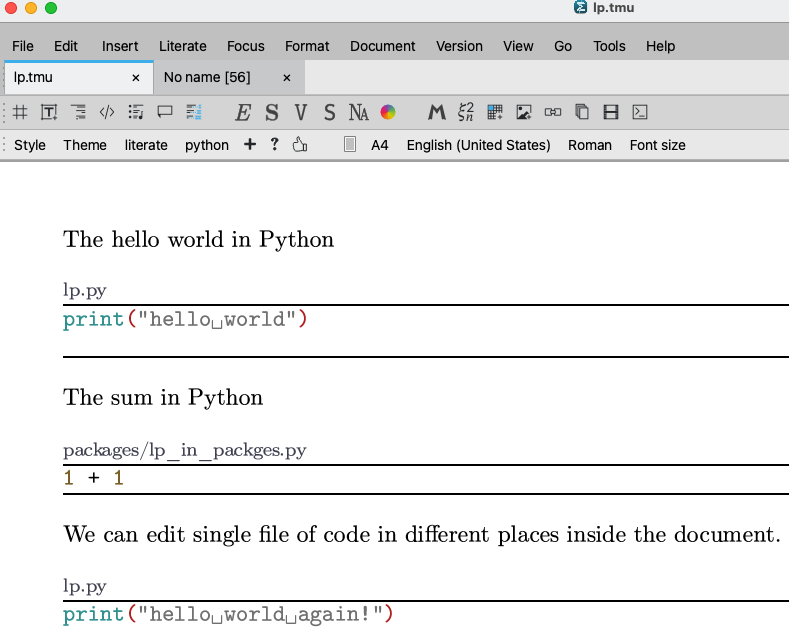}
      \caption{Two files \texttt{packages/lp\_in\_packages.py} and \texttt{lp.py} are edited in a single document. Also, a file \texttt{lp\_in\_packages.py} can be edited at multiple locations within a single document with other files.}
      \label{subfig:mogan_structure}
   \end{subfigure}
   \caption{Comparison: literate programming with Mogan and Jupyter Notebook programming}
   \label{fig:compare_lp_with_jupyter}
\end{figure*}

\begin{figure}[ht]
   \centering
   \includegraphics[width=0.8\linewidth]{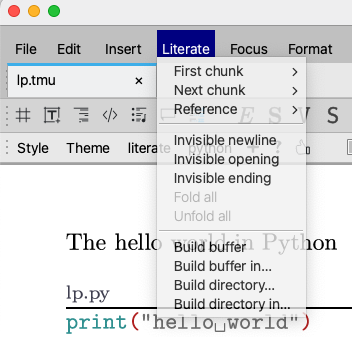}
   \caption{The whole project can be exported from the literate programming document with \textsc{Build buffer} button in Mogan.}
   \label{fig:mogan_project_export}
\end{figure}

\begin{figure*}[ht]
   \centering
   \begin{subfigure}[b]{0.45\textwidth}
      \centering
      \includegraphics[width=\textwidth]{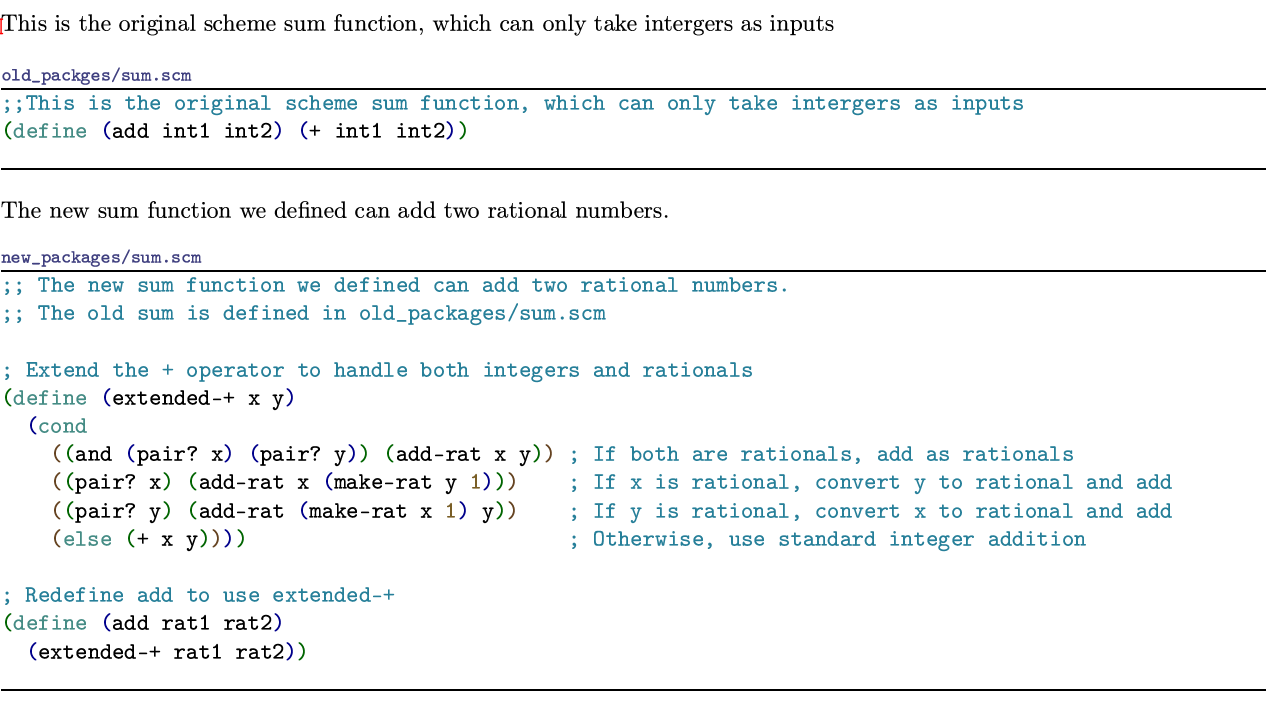}
      \caption{Two idea-connected files edited in Mogan using literate programming; \ac{llm}s can directly read the connection.}
      \label{subfig:web_idea_mogan}
   \end{subfigure}
   \hfill
   \begin{subfigure}[b]{0.5\textwidth}
      \centering
      \includegraphics[width=\textwidth]{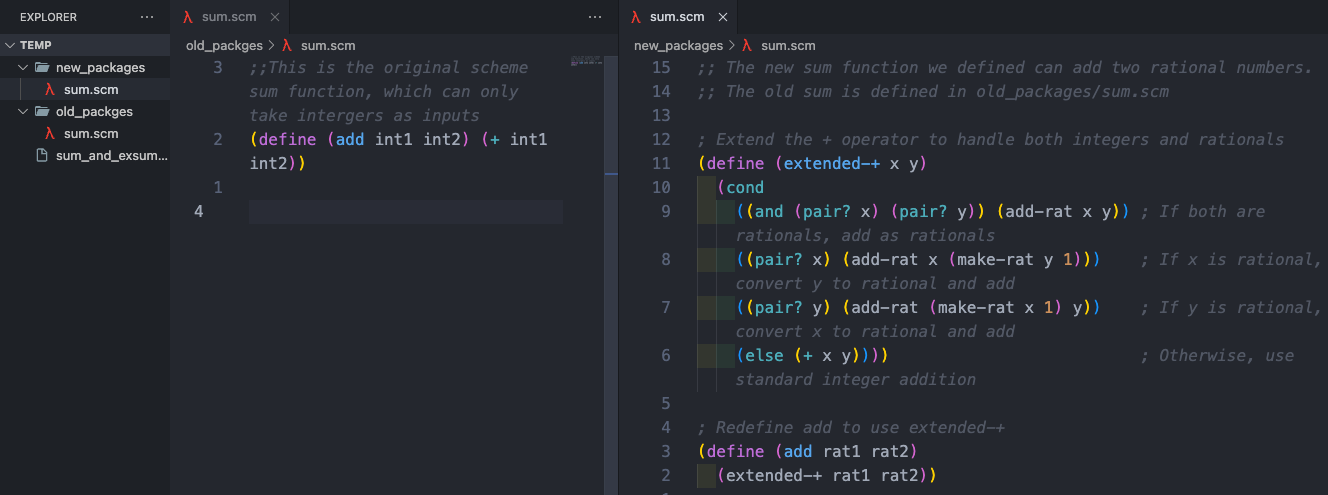}
      \caption{Connection between these two files are commented in \texttt{new\_packages/sum.scm}. Even with such comments, \ac{llm}s must read and interpret the paths mentioned within, locate the referenced file, and then analyze it to grasp the connection.}
      \label{subfig:web_idea_vscode}
   \end{subfigure}
   \caption{Comparison of a literate programming approach (left), which organizes code as a “web” of interconnected ideas, versus a typical Jupyter Notebook workflow (right) that does not follow literate programming principles.}
   \label{fig:lp_flexible_web_of_idea}
\end{figure*}

\subsection{Challenges in LLM-based code generation} \label{subsec: challenges in LLM code generation}
\ac{llms} are generative models designed for natural language processing tasks, capable of performing functions such as text generation, translation, code generation, sentiment analysis, and more.  More discussion on \ac{llm}-based code generation is attached in Sec.~\ref{app:llm-based-code generation}.

Despite their capabilities, \ac{llm}s face limitations in code generation for large-scale projects. In this paper, we focus on the following challenges:

\begin{itemize}
   \item \textbf{External bias}: \ac{llm}s often rely on pre-trained knowledge and external references rather than utilizing the APIs provided within a project. For instance, when tasked with implementing a backpropagation algorithm using a project's APIs, \ac{llm}s are likely to bypass the given APIs and instead default to those in frameworks like PyTorch. Sec.~\ref{app:example-of-external-bias} provides a practical example of the external bias. 
   \item \textbf{Inconsistent Solutions}: \ac{llm}s are highly sensitive to input prompts or context. For instance, changing an irrelevant token in the input can result in entirely different outputs. This inconsistency can lead to unreliable results, especially when stable and predictable solutions are essential for tasks involving complex logic or project-specific APIs.
\end{itemize}

The primary reasons why \ac{llm}s suffer from external bias and inconsistent solutions are the abundance of training data for languages like Python and their widespread popularity. These factors result in an overwhelming number of pre-existing tools and libraries (``wheels'') for \ac{llm}s to choose from, leading to variability in the solutions generated during each run. Additionally, relying on wheels obscures the logical flow of the project. As a result, it becomes challenging for \ac{llm}s to deduce the project's internal logic based solely on the provided codebase.

To enhance \ac{llm}-based code generation in large-scale projects, we reintroduce \ac{lp} (Sec.~\ref{subsec:intro_ILP}) alongside the Scheme language (Sec.~\ref{sec:goldish}). \ac{lp} organizes code and descriptive content in a logical manner, enabling \ac{llm}s to better comprehend the inherent logical flow of the project.

Additionally, we extend the concept of \ac{lp} into \ac{ilp} in Sec.~\ref{subsec:intro_ILP}, where the inherent logical flows between APIs are represented using a directed acyclic graph (DAG) structure within the \ac{lp} document. In this paper, we use Scheme for these descriptions due to its simplicity and inherent $\lambda$-calculus foundation, which naturally aligns with the DAG representation.
Optionally, Scheme can also be used as the target language, enabling \ac{llm}s to generate more consistent code without external bias, as Scheme lacks the extensive libraries and 'wheels' commonly associated with more popular programming languages.
%
The detailed discussion of why the Scheme is suited for the \ac{ilp} documents is in Sec.~\ref{app:scheme-language}.

\subsection{Interoperable \ac{lp}}\label{subsec:intro_ILP}

There is no formal governing body or ISO-style standard for \ac{lp}. There are only several widely agreed-upon practices. The closest to a ``standard'' we can find is Knuth's original WEB system introduced in 1984, which established some foundational concepts:
\begin{itemize}
   \item \textbf{Flexible code organization:} Code can be arranged in any logical order that suits the narrative, interwoven with descriptive text rather than mere comments.\footnote{Figure~\ref{fig:lp_flexible_web_of_idea} illustrates how descriptive text differs from standard comments.} This entire collection of code and text resides in a single document, as shown in Figure~\ref{fig:lp_flexible_web_of_idea}.
   \item \textbf{Tangling:} The combined code must be “tangled” to form a compilable source. While Jupyter does not support this process, Mogan includes tangling functionality, as seen in Figure~\ref{fig:mogan_project_export}.
   \item \textbf{Bidirectional workflows:} Tools and frameworks for \ac{lp} often allow generating documentation from code (weaving) and extracting code from documentation (tangling) within the same environment. This “bidirectional” capability ensures that changes in the narrative are reflected in the compiled code and vice versa.
\end{itemize}
The focus of his design is on making programs comprehensible to humans first and executable by computers second. LP is particularly effective for algorithm-intensive, complex applications, such as large projects, where natural language enhances programmers' understanding of the code. 
LP allows users to edit and export all project files within a single document, aligning with human logic, as illustrated in Fig.~\ref{fig:compare_lp_with_jupyter} and Fig.~\ref{fig:mogan_project_export}. 
Additionally, as shown in Fig.~\ref{fig:lp_flexible_web_of_idea}, LP facilitates connections between files. In contrast, in traditional methods, the logical flow between APIs is inherently hidden throughout the project files.
LP overcomes this limitation by utilizing natural language to establish these connections.

In this paper, we use the term \emph{documents} or \emph{documentations} to refer to scripts that integrate code and descriptive content in a style consistent with \ac{lp} principles.
The term \emph{files} denotes scripts consisting of code and comments. Documents in \ac{lp} support a sequence that feels natural and coherent, where each ``\emph{chunk}'' of code contributes to an unfolding narrative, making the program easier to understand and compile.

Some engineers considered the Jupyter Notebook as an LP document.
However, we do not consider a \texttt{.ipynb} file to be a \emph{document} but rather a \emph{file} in this paper.  While a Notebook can mix text and code, it is not easily integrated or exported into a broader project structure in the way that \ac{lp} requires. This lack of portability separates Notebook-based code from the type of unified narrative and codebase advocated by \ac{lp} in this paper. Additionally, it does not allow files of codes to be structured as a flexible ``web'' of ideas in a single document. A more detailed discussion can be found in Sec.~\ref{app:jupyter and LP}. It is also important to note that code with detailed comments does not constitute \ac{lp}; this topic is further explored in Sec.~\ref{app:jupyter and LP}.

To distinguish our definition of \ac{lp} from Notebook-based coding, we introduce a modern form of \ac{lp} called \acf{ilp}. \ac{ilp} incorporates all \ac{lp} features we listed at the beginning of this subsection. More discussion on \ac{ilp}is in Sec.~\ref{app:jupyter and LP}.

\subsection{Goldfish Scheme as nature language compensation for descriptive contents in \ac{ilp}}\label{sec:goldish}

In this work, we introduce \emph{Goldfish Scheme}, a Scheme interpreter with a Python-like standard library. As discussed in Sec.~\ref{subsec: challenges in LLM code generation}, Scheme is used to describe a directed acyclic graph (DAG) that represents the inherent logical flow between APIs in Python projects, leveraging Scheme's simplicity and its $\lambda$-calculus foundation.
Goldfish Scheme incorporates these features and includes an LLM interaction layer, making it highly suitable for LLM-based code generation. Additional features, such as its Python-like standard library for transfer learning, are further discussed in Sec.~\ref{appendix:goldfish scheme}.

\paragraph{\ac{llm} interaction layer}

The core idea behind this interaction layer is to supply machine-readable fields that describe key aspects of a function or library component. These annotations can specify algorithmic patterns, complexity metrics, stability characteristics, and concrete usage examples. By making this information explicit and accessible, the interaction layer allows \ac{llm}s to understand each piece of code not only at a syntactic level but also in terms of conceptual intent and practical implications.
Consider the following \texttt{define-with-docs} macro:

\begin{figure}
    \centering
    \begin{minted}[breaklines, fontsize=\small]{scheme}
    (define-with-docs quicksort
      #:pattern "divide-and-conquer"
      #:complexity "O(n log n)"
      #:stability "unstable"
      #:examples
      '((quicksort '(3 1 4 1 5 9 2 6 5 3))
        => (1 1 2 3 3 4 5 5 6 9))
      (lambda (lst)
        ;; Implementation follows...
        ))
    \end{minted}
    \caption{Goldfish Scheme as descriptive contents in \ac{ilp}.}
    \label{fig:define-with-doc}
\end{figure}

Remember that the macro is not intended for execution but serves as an alternative to natural language in the descriptive content of an \ac{ilp} document. The goal is not to generate or execute Scheme code but to facilitate the generation of Python code.
In Fig.~\ref{fig:define-with-doc}, the ``divide-and-conquer`` pattern informs the \ac{llm} of the underlying strategy behind the algorithm, making it easier for the model to relate this sorting routine to other similar functions and patterns it knows. By stating that the sort is ``unstable`` and complexity as \texttt{O(n log n)}, the annotations reveal a critical property that distinguishes the function’s behavior, guiding the \ac{llm} toward more precise reasoning about potential use-cases and limitations. The inclusion of example tags allows the \ac{llm} to confirm its understanding of inputs, outputs, and expected transformations. Such an example can also serve as a reference point for generating further tests or variant solutions. More discussions are in Sec.~\ref{appendix:goldfish scheme}.

%% file: sections/4_our_work.tex
\section{Interoperable literate programming documentation design} \label{sec:ilp}

In Sec.~\ref{subsec: challenges in LLM code generation}, we identified multiple challenges in using LLMs for code generation in large-scale projects. We propose addressing these challenges by applying \ac{ilp} with descriptive content that structures an API’s logic as a Directed Acyclic Graph (DAG). In this section, we present our methodology for creating the \ac{ilp} document.
Although we use Scheme code snippets to illustrate how the DAG is documented, these snippets are not intended for execution or compilation. Instead, they serve as placeholders to capture the logical flow of an API, allowing us to design and describe interdependencies without running the code itself.


A clear and logically structured documentation design is crucial for explaining API logic and guiding LLMs in generating correct code. We propose a step-wise framework to tackle the challenges from Sec.~\ref{subsec: challenges in LLM code generation}. This framework reduces reliance on external references, enforces logical consistency, and lowers the risk of errors.

We adopt a step-wise methodology inspired by mathematical induction and recursive algorithms, dividing large tasks into base (\emph{zero-step}) and inductive (\emph{successor-step}) parts. For instance, in recursive code, establishing a clear base case and carefully explaining the recursive step ensures consistency in the final output. Similarly, our method segments a task’s logic into smaller, sequential units, giving LLMs an incremental path instead of overwhelming them with entire program logic at once. By breaking down complex tasks into separate components, each step can be understood independently. Additional benefits are discussed in Sec.~\ref{app:benefit_of_step_wise}. This structure inherently forms a DAG.
Our approach relies on two core principles: \emph{Scheme’s simplicity} and a \emph{stepwise progression}. Scheme’s minimal syntax and functional style treat each part of the logic as a discrete unit. Furthermore, because Scheme is relatively uncommon in LLM training data, models must depend on the provided documentation, reducing the potential for errors caused by external biases (see Sec.~\ref{subsec: challenges in LLM code generation}).

%
%

Below, we outline the key components that enable this method to function effectively (additional details appear in Sec.~\ref{app:benefit_of_step_wise}):
\begin{enumerate}
    \item \textbf{Zero-step definition}: We start by specifying a \emph{base case} or fundamental building block. In recursive functions, for example, this might be the condition under which recursion terminates. By designating this zero-step, we provide the LLM with a clear starting point for the broader logic.
    
    \item \textbf{Successor-step explanation}: We then describe how one logical state transitions to the next. In inductive or recursive algorithms, this step outlines the ``successor'' or ``next element,'' ensuring that each additional step builds consistently on the zero-step.
\end{enumerate}

To clarify these concepts, we offer several toy examples that target the Scheme language and use the R7RS document as a base for literate programming. We embed the following descriptive content in the R7RS document, converting it into an \ac{ilp} document:
\begin{minted}[breaklines, fontsize=\small]{markdown}

## take-right

### Zero-Step Logic
Return the list if it's empty.

### Succ-Step Logic
See below;

### Helper Function: `drop`

The helper function `drop` removes the first *n* elements from a list. See Sec. 7.1.7 in this document.
\end{minted}
\begin{minted}[breaklines, fontsize=\small]{scheme}
(define-with-docs take-right
  #:pattern "divide-and-conquer"
  #:complexity "O(n)"
  #:stability "stable"
  #:examples
  '(take-right '(a b c d e) 2) => (d e)
  (lambda (lst)
    ;; Implementation follows...
    ;; Base (zero-step):
    ;;   if n <= 0 or lst is empty, return lst as is
    ;; Succ-step:
    ;;   otherwise, (drop lst 1) and (take-right lst (- n 1))
    ))
\end{minted}

We then prompt LLMs with:
\begin{quote}
    Write the function (take-right flist i) using the zero-step/succ-step approach as described in the pdf file. The Scheme language grammar and necessary APIs can be found in the /opt/cachebox/r7rs/.
\end{quote}

\begin{figure*}[ht]
   \centering
   \begin{subfigure}[b]{0.45\textwidth}
      \centering
      \begin{minted}[breaklines, fontsize=\small]{scheme}
;; The following are documents in Chapter. 3
(define-with-docs add
  #:pattern "api-calls"
  #:complexity "undefined"
  #:stability "stable"
  #:examples
  '(add (make-rat 1 2) (make-rat 1 2))=> 1
    ;; Implementation follows...
    ;; Base (zero-step):
    ;;   + in original Scheme
    ;; Succ-step:
    ;;   call extended-+ in Chapter.2 
    ))
;; The following are documents in Chapter. 2, searched by LLMs.
(define-with-docs extended-+
  #:pattern "api-calls"
  #:complexity "undefined"
  #:stability "stable"
  #:examples
  '(extended-+ (make-rat 1 2) (make-rat 1 2))=> 1
    ;; Implementation follows...
    ;; Base (zero-step):
    ;;   + in the original Scheme
    ;; Succ-step:
    ;;  call add-rat and make-rat in Chapter. 2. 
    ;;  call pairs? in Chapter 6.
    ))
;; more define-with-docs follows
\end{minted}
      \caption{Scheme representation for DAG: Note that these representations are located in different chapters of the \ac{ilp} document. Only \texttt{(define-with-doc add)} needs to be written, as LLMs can recursively search for other descriptions previously written. Hyperlinks are included to reference the related chapters within the document.}
      \label{subfig:scheme_description_for_extended_add}
   \end{subfigure}
   \hfill
   \begin{subfigure}[b]{0.5\textwidth}
      \centering
      \includegraphics[width=\textwidth]{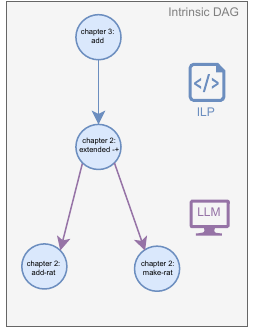}
      \caption{DAG representation: Each node corresponds to a Scheme description, with nodes distributed across different chapters. As shown in purple lines, LLMs shall trace the inherent logical flow of APIs through the descriptive contents in the document.}
      \label{subfig:dag_for_exteded_add}
   \end{subfigure}
   \caption{The DAG structure and the corresponding Scheme description of the extended \texttt{add} function are shown in Fig.~\ref{subfig:web_idea_mogan}.}
   \label{fig:example_of_scheme_dag_description}
\end{figure*}

All toy-case experimental results appear in Sec.~\ref{app:toy_result} and Table~\ref{tab:function-success}. As shown in Table~\ref{tab:function-success}, GPT-4o generates correct code for all toy examples when guided by \ac{ilp}. 
In the toy-case example above, the zero-step and successor-step align closely with the recursive implementation of \texttt{take-right}. However, not all functions rely on pure recursion. In some instances, the zero-step and successor-step capture the broader logical flow of APIs in the \ac{ilp} document rather than a direct mapping to implementation details. Fig.~\ref{fig:example_of_scheme_dag_description} illustrates the DAG structure and corresponding Scheme description for the extended \texttt{add} function from Fig.~\ref{subfig:web_idea_mogan}. Another example involving Python code generation from a Scheme-based description is provided in Sec.~\ref{app:python-api-design-in-scheme}.

Finally, the stepwise method itself is inherently recursive, and any code generated by LLMs reflects this pattern. Adding the generated code segments back into the \ac{ilp} document produces an updated \ac{ilp} document, reinforcing the same iterative design.

%% file: sections/5_exp.tex
\section{Experiments} \label{sec:exp}

\input{tables/scheme-code-generation}

In this section, we designed our experiments to observe the performance of LLMs' Python code generation by learning ILP-style documents. Our experiments are conducted on two types of LLMs: local open-source LLMs and remote commercial LLMs. The machine details are in Sec.~\ref{app:machine}.





\subsection{RepoBench: benchmark for code generation} \label{subsubsec:RepoBench_justification}

RepoBench~\citep{liu_repobench_2023} serves as our evaluation benchmark environment due to its realistic and comprehensive testing framework. Unlike benchmarks focused on isolated coding tasks, RepoBench emphasizes the interconnection between code components within projects. This design provides a practical environment for evaluating how LLMs handle dependencies and relationships across different parts of a codebase, reflecting the complexities of real-world software development.

RepoBench's test cases are sourced from GitHub repositories, ensuring authenticity and diversity in the scenarios it presents. These real-world examples go beyond simplified toy problems, capturing the nuances and challenges typical of production-level software. This allows us to evaluate the performance of LLMs in contexts that closely mirror actual programming tasks.

Another critical feature of RepoBench is its diverse documentation patterns, which include comments in various programming languages. While most comments are not originally written in Scheme, this diversity presents a valuable opportunity to test the adaptability of our approach. To enable Scheme-based evaluations, we translated a selection of comments into Scheme, preserving their semantic meaning while aligning them with Scheme's functional programming paradigm.

Further details about our adaptation of RepoBench for Scheme are discussed in Sec.~\ref{subsec:repobench for scheme}. In Sec.~\ref{subsec:whyotherworesethan repobench}, we provide a thorough comparison with alternative code generation benchmarks, highlighting why RepoBench is particularly well-suited for evaluating LLM-based code generation in our study.

\subsection{Experiment: Why Scheme is ideal for LLM code generation?}

In Sec.~\ref{sec:goldish} and Sec.~\ref{sec:whyscheme}, we discussed why Scheme is ideal as a description language for LLM code generation. The main reasons are: 1. LLMs are not extensively trained in Scheme. 2. Scheme is better suited for describing logical structures. In this section, we design an experiment to verify these conjectures.

We first examined the extent of Scheme knowledge embedded in LLMs by prompting ChatGPT-4O with basic requests like ``Generate a function called \texttt{xxx} in Scheme that outputs \texttt{xxx}.''
The resulting code often contained syntax errors or flawed logic, suggesting that the models do not possess significant pre-trained Scheme patterns. Fig.~\ref{fig:sort-int-llm-fail} in Sec.~\ref{app:scheme-language} provides an example where ChatGPT-4 fails to generate a simple \texttt{sort-int} function. 
This finding supports our hypothesis that Scheme's limited presence in modern development contexts forces LLMs to engage in actual logical reasoning rather than recalling ready-made examples.

Next, we investigate whether Scheme's mathematical underpinnings enhance logical reasoning. We compare code generation using standard \href{https://r7rs.org/}{R7RS} and the \ac{ilp} document based on R7RS discussed in Sec.~\ref{sec:ilp}.

As shown in Table~\ref{tab:function-success}, we test operations ranging from basic list manipulations to more advanced tasks like \texttt{delete-duplicates} and \texttt{split-at}. When comparing different prompts—R7RS-only descriptions, R7RS with detailed documentation, and R7RS with instructions to ``fully depend on the file''—we observed that performance improved significantly with the \ac{ilp} documentation. Even when we replaced standard function names with arbitrary terms, the models generated working code, implying that they were guided by the documented logic rather than default training data.

What's more, to ensure LLMs focus solely on our documentation rather than their broader training knowledge, we implement specific constraints in our prompts. One key strategy involves using non-standard function names in our implementation. Instead of traditional function names like \texttt{map}, we experiment with arbitrary names, such as Pinyin (Mandarin Chinese that uses the Latin alphabet) translations. This approach helps us verify whether LLMs are truly working from our documentation rather than falling back on their pre-trained knowledge of the R7RS Scheme.

For example, rather than using:
\begin{minted}[breaklines, fontsize=\small]{scheme}
(define (map function lst)
  ...)
\end{minted}
We might use:
\begin{minted}[breaklines, fontsize=\small]{scheme}
(define (ditu hanshu liebiao)
  ...)
\end{minted}

This naming strategy serves as a clear indicator: if an LLM uses the standard name in English instead of our arbitrary one, we know it's drawing from external knowledge rather than our documentation. The result is given in Sec.~\ref{sec:experimental-details}.

These findings confirm that Scheme's $\lambda$-calculus structure and reduced prevalence in training sets encourage LLMs to rely on logical reasoning. Furthermore, our two-step documentation pattern, based on mathematical induction, effectively channels LLM reasoning by splitting code generation into clear base and inductive steps. This design takes advantage of Scheme's alignment with theoretical principles, leading to reliable code generation for a broad range of tasks.

\subsection{The general design of a single \ac{ilp} document}

To improve LLM performance in Goldfish Scheme, we structure the natural language portion of our documentation using the \ac{cot} methodology, enhancing LLMs' understanding of code relationships. The other descriptive contents follow the stepwise method outlined in Sec.~\ref{sec:ilp}. Code chunks follow directly with the corresponding descriptive contents. More discussions are in Sec.~\ref{sec:experimental-details}.

\subsection{Inter-file inference management}

We carefully structure our code to test LLMs' ability to follow logical connections across files. Instead of maintaining traditional linear dependencies (where A depends on B, which depends on C), we split implementations across files in ways that require an understanding of the broader context.

Consider a typical implementation flow:
\begin{enumerate}
  \item Input processing in File A.
  \item Data transformation in File B.
  \item Resulting LLM-generated code chunks in File A.
\end{enumerate}

This structure tests whether LLMs can:
\begin{itemize}
  \item Recognize the overall logical flow.
  \item Identify connections between seemingly separate implementations.
  \item Understand how different components work together despite being physically separated.
\end{itemize}

Fig.~\ref{fig:example_of_scheme_dag_description} shows a case for these type of code generation tasks. The following is another example of this structure:

\begin{figure}[ht]
  \centering
  \begin{minted}[breaklines, fontsize=\small]{scheme}
    ;; File: processing.scm
    (define (stage-one input)
      (process-initial input))
    
    (define (stage-three processed-data)
      (generate-result processed-data))
    
    ;; File: transform.scm
    (define (stage-two data)
      (transform-intermediate data))
\end{minted}
  \caption{Cross-file implementation example}
  \label{fig:implementation}
\end{figure}

This structure helps us evaluate not just whether LLMs can write correct code but whether they truly understand the relationships and dependencies within our codebase. The non-linear organization of functionality across files serves as a more rigorous test of an LLM's comprehension of the overall system architecture.

\subsection{Experiment result on Python code generation on RepoBench} \label{subsec:exp-python-code-repobench}

Take the generation of the \texttt{take-right} function on RepoBench as an example; the \ac{ilp} document is the same as in Sec.~\ref{sec:ilp} and the prompt given to LLMs are:
\begin{quote}
    Fully based on the file, generate a function in python for take-right API mentioned in the document?
\end{quote}

We test code generated for the \texttt{take-right} function, both with and without ILP-based guidance. The experiment result of Python code generation on RepoBench is in Sec.\ref{app:repobench_result}. The result reveals measurable improvements in structural consistency and implementation quality with the help of \ac{ilp}.

When following ILP documentation, large and small models alike produce similar code structures. GPT-4, Claude Sonnet, and LLAMA 3.1 2B implementations all include well-defined edge case handling, helper functions such as \texttt{drop}, and comprehensive error checks. In contrast, implementations generated without ILP guidance vary significantly in structure; some omit essential edge case handling and error checks altogether.

Local LLMs with fewer parameters show marked gains under ILP guidance. For example, LLAMA 3.1 2B generates code that includes proper error handling and documentation, matching the functional characteristics of larger models like GPT-4. This indicates that ILP’s structured documentation method enables smaller models to produce code of comparable reliability.

All tested models, including both commercial online systems (GPT-4 and Claude) and local deployments (LLAMA and Qwen), exhibit consistent and systematic error-handling patterns under ILP. Each implementation checks for \texttt{null?} conditions and boundary cases while preserving functional programming practices such as avoiding side effects and implementing correct recursion. 

ILP’s structural elements directly support these improvements. The zero-step/successor-step approach aligns with recursive design, clarifying base cases and recursive transitions. The documentation specifies error conditions and edge cases, providing clear patterns for helper function implementation. By incorporating functional programming principles directly into the documentation, ILP guides models toward functional purity and correct recursion in their implementations.

%% file: tables/scheme-code-generation.tex
\begin{table*}[ht]
  \centering
  \caption{Scheme Function Implementation: Success or Failure. ``R.'' refers to the detailed descriptions from R7RS, while ``D.'' represents the designed documentation following our two-step programming logic.}
  \label{tab:function-success}
  \resizebox{\textwidth}{!}{%
    \begin{tabular}{l p{8cm} c c c}
      \hline
      Function          & Description                                                                                                                & GPT-4o (R. D.) & GPT-4 (R. D.) & GPT-4o (R.) \\
      \hline
      make-list         & Returns an \texttt{n}-element list, whose elements are all the value \texttt{fill}.                                        & S              & S             & S           \\
      iota              & Returns a list containing the elements (\texttt{start}, \texttt{start+step}, ...~, \texttt{start+(count-1)*step}).         & S              & S             & F           \\
      null-list?        & Returns true if it is the empty list \texttt{()}, and false otherwise.                                                     & S              & S             & S           \\
      drop-right        & Returns all but the last \texttt{i} elements of the given list.                                                            & S              & S             & S           \\
      take-right        & Returns the last \texttt{i} elements of the list.                                                                          & S              & S             & S           \\
      split-at          & Splits the list \texttt{x} at index \texttt{i}, returning a list of the first \texttt{i} elements, and the remaining tail. & S              & S             & F           \\
      append            & Returns a list followed by another list.                                                                                   & S              & F             & S           \\
      reverse           & Returns a newly allocated list consisting of the elements of the list in reverse order.                                    & S              & S             & S           \\
      for-each          & Calls \texttt{proc} (given functions) on the elements of the lists.                                                        & S              & S             & S           \\
      remove            & Returns list without the elements that satisfy predicate \texttt{pred}.                                                    & S              & S             & S           \\
      find              & Returns the first element of \texttt{clist} that satisfies predicate \texttt{pred}; false if no element does.              & S              & S             & S           \\
      delete-duplicates & Removes duplicate elements from the list argument.                                                                         & S              & S             & F           \\
      \hline
    \end{tabular}%
  }
\end{table*}

%% file: sections/6_conclusion.tex
\section{Conclusion} \label{sec:conclusion}

This paper introduced a novel framework, \ac{ilp}, which extends the principles of \ac{lp} to address the challenges of large-scale software development with LLMs. By providing a structured approach to managing interdependencies, ensuring consistency, and facilitating efficient code generation and documentation, \ac{ilp} overcomes the limitations of traditional prompt-based methods. Through experiments conducted on RepoBench and practical applications demonstrated with tools like Mogan and Goldfish Scheme, we showed that \ac{ilp} enhances code generation quality, efficiency, and scalability. These advancements make \ac{ilp} a robust solution for modern software projects, enabling developers to seamlessly integrate documentation and achieve reliable outcomes in increasingly complex coding environments. Our findings underline \ac{ilp}’s potential to transform LLM-based development workflows and pave the way for more efficient and coherent project management practices.

%% file: sections/appendix.tex
\input{appendix/jargon}

\input{appendix/llm_based_code_generation}

\input{appendix/ilp_and_scheme}

\input{appendix/goldfish_scheme}

\input{appendix/discuss_step_wise}

\input{appendix/repobench}

\input{appendix/experiment_details}

\input{appendix/lp_in_modern_develop}

\input{appendix/exp_result}

%% file: appendix/jargon.tex
\section{Jargon} \label{app:jargon}

We list commonly used jargon in the following:
\begin{itemize}
  \item \emph{File}: A script containing code and comments without the structured integration needed for \ac{lp} (e.g., Jupyter \texttt{.ipynb} files)
  \item \emph{Document}: A script that combines code and descriptive content in a cohesive, narrative style, following \ac{lp} principles (e.g., Mogan \texttt{.tmu} documents). Unlike in a file, sections of a single file can be edited in multiple locations within a document, with other files interspersed in between.
  \item \emph{Chunk}: A segment of code and comments within a file. Multiple chunks from different files can be organized within a single document.
\end{itemize}

%% file: appendix/llm_based_code_generation.tex
\section{LLM-based code generation} \label{app:llm-based-code generation}

The development of tools based on LLMs, such as \href{https://chatgpt.com/}{ChatGPT},
\href{https://github.com/features/copilot}{GitHub Copilot},
\href{https://openai.com/index/openai-codex/}{OpenAI Codex}, and
\href{https://aws.amazon.com/q/developer/?gclid=CjwKCAiAyJS7BhBiEiwAyS9uNeHuUJCRJrLQbxcSsxm-6pm-L7kGBFYRQZlFb5t2iSM7hTTHP2ARyRoC5bMQAvD_BwE&trk=c570e8a2-ec3c-4968-baa4-f8537e37dd1d&sc_channel=ps&ef_id=CjwKCAiAyJS7BhBiEiwAyS9uNeHuUJCRJrLQbxcSsxm-6pm-L7kGBFYRQZlFb5t2iSM7hTTHP2ARyRoC5bMQAvD_BwE:G:s&s_kwcid=AL!4422!3!698234556950!e!!g!!amazon%20codewhisperer!21048268530!156938945890}{Amazon Q Developer}, has transformed the way programmers solve problems, significantly enhancing efficiency and convenience. These models streamline various coding tasks, including code completion~\citep{guo2023longcoder, lu2022reacc, wang2021code, wu2024repoformer, svyatkovskiy2019pythia, fried2022incoder}, code repair~\citep{parasaram2024fact, zhang2024pydex, fan2023automated, joshi2023repair, olausson2023self, jin2023inferfix, xia2023automated}, code translation~\citep{chen2018tree, roziere2020unsupervised, szafraniec2023code}, and code comprehension~\citep{nam2024using, song2019survey}. 

Despite their impressive capabilities, LLMs still face limitations in certain areas, particularly in code generation, where solution inconsistency and specification complexity pose two fundamental challenges. These challenges hinder the development of reliable, scalable software systems through LLM assistance.
For instance, \citet{roziere2020lost} highlights the challenges of using LLMs for code translation, demonstrating that these models can introduce bugs during the translation process, especially when dealing with complex or domain-specific code. \citet{dou2024s} highlights the limitations of current LLMs in handling large-scale software projects, where the scale of the project and the interdependencies within the codebase create significant challenges. The study finds that LLMs often struggle to generate functional code for larger-scale projects, and their accuracy decreases as the complexity of these interdependencies increases. Such limitations emphasize the need for more robust frameworks to ensure reliability in large-scale code-related tasks.

\subsection{Chanllenges of LLM-based code generation}
In the following, we list several common challenges for LLM-based code generation:

\paragraph{Solution inconsistency}
Solution inconsistency refers to LLMs’ tendency to generate varied implementations for the same problem specification. This inconsistency can manifest in different forms, such as variations in computational complexity, coding style, or error-handling strategies. For example, a list processing task might have an efficient $O(n)$ implementation using hash tables. However, given a high-level description, LLMs often generate less efficient nested-loop solutions with $O(n^2)$  complexity. While both solutions may produce correct results, their significant differences in performance characteristics can create challenges in production systems, where performance and reliability are critical.

\paragraph{API specification complexity}
The complexity of modern API specifications presents significant challenges for LLM-based code generation. APIs often encompass multiple interconnected functionalities that must be implemented cohesively. For example, a database connection API involves configuration validation, connection pooling, retry logic, error handling, and resource cleanup. However, LLMs often struggle to comprehend the intricate relationships between these components, leading to incomplete or incorrect implementations. This issue is exacerbated by ambiguous or incomplete documentation, which prevents LLMs from fully understanding the API's intended behavior.

\paragraph{Inherent LLM limitations}
LLMs’ inherent limitations, such as fixed context windows and constrained processing time, further compound the challenges of generating accurate code. For example, when dealing with complex API specifications or intricate business logic, the model may truncate or omit essential information, resulting in incomplete code. A common scenario involves error handling logic being omitted when the context exceeds the model's capacity. These limitations necessitate significant manual intervention, reducing the efficiency of LLM-based code generation in complex environments.


\subsection{An example of external bias} \label{app:example-of-external-bias}

The following shows an example of the external bias:

\begin{itemize}
  \item Input:

        Question: Implement the "filter" in [[goldfish]]. The "filter" [[procedire]] receives a [[List]] named "lst" and a boolean [[procedure]] "fn" as [[input]]; it checks every element in the "lst" and collects those who have TRUE result by "fn".

        Thinking: I lookup the document of [[goldfish]] carefully. I will learn how to write a [[procedure]] on [[List]] according to the [[induction]] method. I will foucs on the document.
        I write it in the steps described by [[induction]].

        Answer:

  \item Output:
        \begin{minted}[breaklines, fontsize=\small]{scheme}
        (define (filter lst fn)
          (define inner-filter
            (lambda (result lst)
              (;zero-step (if (null? lst) result (<succ-step>))))))
          (inner-filter lst 0))
        \end{minted}

        One can see that the result build the "Skeleton of Chunk" correctly. However, the generation of "<succ-step>" differs.

\end{itemize}

%% file: appendix/ilp_and_scheme.tex
\section{Discussion of literate programming and Scheme language}

\subsection{ILP, Jupyter Notebook, and code with detailed comments} \label{app:jupyter and LP}

Fig.~\ref{fig:compare_lp_with_jupyter} illustrates the differences between Python \ac{lp} in Mogan and Python programming in Jupyter Notebook.
As shown in Figure~\ref{subfig:jupyter}, two separate \texttt{.ipynb} files contain code and markdown for different parts of the same project. When using Jupyter, each notebook typically focuses on a single task or dataset, making it difficult to combine multiple code sections or logic threads into one unified file. In contrast, as shown in Mogan (Figure~\ref{subfig:mogan_structure}), both code files can be edited in a single \texttt{.tmu} document.
Furthermore, Fig.~\ref{subfig:mogan_structure} shows three chunks belonging to two different files, all within one document.
Once editing is complete, Mogan allows users to export all project files at once, as depicted in Fig.~\ref{fig:mogan_project_export}.
For the same reason discussed above, files with detailed comments, such as the Natural Language Outlines introduced by~\citet{shi_natural_2024}\footnote{As noted by~\citet{shi_natural_2024}, the Natural Language Outlines lack a utility to export all project files cohesively. This limits alignment with the \ac{lp} philosophy discussed in Section J of~\citet{knuth_literate_1984}.} are not considered documents.

We now show another example of the difference between code with detailed comments and \ac{lp}.
As discussed in Section J of \citet{knuth_literate_1984}, \ac{lp} enables a flexible structure where code is organized as a "web" of interconnected ideas rather than following a strict top-down or bottom-up sequence.
Fig.~\ref{fig:lp_flexible_web_of_idea} illustrates this concept with two files connected by ideas. The file \texttt{old\_packages/sum.scm} contains the original sum function in Scheme, which only accepts integer inputs.
The file \texttt{new\_packages/sum.scm} extends this sum function to accept rational numbers. These two files are linked by ideas rather than function calls or class inheritance.
Fig.~\ref{subfig:web_idea_mogan} demonstrates how \ac{llm}s can directly recognize these connections, as literate programming allows these chunks to be organized this way.
In contrast, Fig.~\ref{subfig:web_idea_vscode} shows the same project with detailed comments. Even with such comments, \ac{llm}s must read and interpret the paths mentioned within, locate the referenced file, and then analyze it to grasp the connection.

To distinguish our definition of LP from others,  we introduce a modern form of \ac{lp} called \emph{Interoperable Literate Programming} (ILP). ILP addresses the persistent fragmentation and inconsistency in today's \ac{lp} practices. Existing approaches rely on tool-specific formats, ad-hoc documentation styles, and inconsistent code extraction methods. These conditions make it difficult to transfer established practices, incorporate new technologies, or maintain compatibility over time. ILP offers a stable, language-agnostic foundation that can adapt to various workflows, tools, and programming languages.

ILP defines a uniform, human-readable markup base that integrates code and documentation. Instead of embedding every feature at a single level, ILP specifies a minimal core schema governing how code, narrative, and metadata coexist. Within this schema, code blocks and their annotations—covering complexity, algorithmic patterns, and related functions—follow a consistent structure. This approach ensures that compilers, documentation generators, analysis tools, and editors can parse and process ILP documents without ambiguity.

Building on its core specification, ILP supports incremental adoption of optional layers. These layers introduce capabilities—such as typed code sections, debugging aids, or hints for large language models—without altering the underlying standard. This arrangement preserves interoperability, since tools that do not require certain features can ignore them, and new layers can emerge without forcing widespread rewrites. Developers, educators, and researchers gain the flexibility to select enhancements that suit their context while relying on a well-defined core.

ILP also encourages formal interoperability contracts that describe how code extraction, documentation, and compilation occur for specific languages or frameworks. By establishing these contracts alongside the base specification, ILP makes it easier to introduce new languages, connect existing codebases, and update tooling. This structure helps avoid the proliferation of incompatible formats and supports incremental refinement as technology evolves.

\subsection{Thinking in interoperable literal programming} \label{app:thinking-in-ilp}

The philosophy of using ILP is not to adopt an entirely new way of thinking but to transfer one's natural thought process into \ac{lp}. Software and software development should aim to provide an intuitive user experience, aligning with the brain's natural thinking patterns, whether sequential or otherwise. Consequently, the impact of software lies not in the new features it offers but in how well it preserves and supports users' original ways of thinking.

Using some cases in prompt engineering as examples, scholars try to understand the intelligent thinking process of LLM; as a result, they provide many different methods. The observation of the \ac{cot} and other prompts engineering methods provides us with an important idea that the continuous working flow is of great importance for a long-lasting task. Indeed, sequential thinking, like the \ac{cot}, and tree-like or more complex patterns need the software to have enough flexibility. Mogan provides writing with code chunks at any place, which is the key to combining thinking with writing. In other words, the whole spectrum of thinking and writing flow in the human brain can be preserved perfectly according to the Literal Programming based on Mogan.


In the following sections, we propose a structured approach for documentation and code organization in Goldfish Scheme, aimed at enhancing LLM performance and understanding. We adopt the classical \ac{cot} pattern to capture both the final code and the thought process behind it, ensuring that documentation 
covers not only what the code does but why specific decisions were made. This involves a functional-programming-inspired template that restricts direct recursion and uses zero-step/succ-step induction. We then extend this strategy with a unique naming system and cross-file references to test whether LLMs rely on our explicit documentation instead of pre-trained knowledge. We can further assess the model's ability to infer relationships and maintain context by splitting functionality across multiple files and introducing dependency loops.

\subsection{Literate programming supported editor}\label{subsec:lp-editor}
\citet{knuth_literate_1984} introduce the concept of \ac{lp} and implement \ac{lp} in the ``WEB'' system, which leads the \TeX program itself, brought to life as a narrative in the book \href{https://ctan.org/tex-archive/info/knuth-pdf}{``\TeX: The Program''}.
Then, in 1994, \citet{ramsey1994literate, johnson1997literate} develop a simplified version,  ``\href{https://www.cs.tufts.edu/~nr/noweb/}{NoWEB}'' which supports not only \TeX but also other text formatting systems, such as \LaTeX and HTML. In addition to ``NoWEB'',  several modern literate programming systems have been developed, such as  ``\href{http://www.ross.net/funnelweb/}{FunnelWEB}'', ``\href{https://sourceforge.net/projects/nuweb/}{NuWEB}'', ``\href{https://ganelson.github.io/inweb/inweb/index.html}{InWEB}'', ``\href{https://slott56.github.io/py-web-tool/src/_build/html/pyweb.html}{PyWEB}'', ``\href{https://orgmode.org/}{Org mode}''~\citep{Dominik2010, Schulte, 5756277}, ``\href{https://savannah.nongnu.org/p/fangle}{Fangle}'' and ``Jupyter Notebook''~\citep{Kluyver2016jupyter} among others. Additionally, {\href{http://www.axiom-developer.org/axiom-website/documentation.html}{Axiom} embodies literate programming, linking code with its rationale for clarity. In this paper, we develop Mogan, a UTF-8-compatible literate programming editor designed to simplify the workflow by allowing users to edit and export all project files within a single document, as illustrated in Fig.~\ref{fig:compare_lp_with_jupyter} and Fig.~\ref{fig:mogan_project_export}.

\subsection{Scheme language}\label{app:scheme-language}

Scheme language is known for its minimal core and uniform syntax. Its straightforward constructs, first-class procedures, and lexical scoping encourage a focus on program logic rather than language intricacies. This simplicity makes Scheme an effective foundation for exploring language design and building upon established standards.

\citet{sussman_scheme_1975} implement ``Scheme'', a simple and lightweight functional programming language based on lambda calculus~\citep{steele1976lambda, steele1976lambdad, steele1977debunking, steele1978art, steele1978rabbit, steele1979design, steele1980compiler}, which becomes widely known through the popular book Structure and Interpretation of Computer Programs (SICP)~\citep{abelson_sicp_1996}. Furthermore, several Scheme interpreters have been developed based on the Scheme standard, with the latest and most widely accepted being \href{https://r7rs.org/}{R7RS}. Additionally, Scheme Requests for Implementation (\href{https://srfi.schemers.org/}{SRFI}) are community-driven specifications that extend the language with additional features and libraries. Among the Scheme interpreters, \href{https://cisco.github.io/ChezScheme/}{Chez Scheme}~\citep{dybvig2005chez, dybvig2009scheme} is a high-performance Scheme interpreter, known for its efficient compiler and extensive features. However, it only supports the R6RS standard. Several other implementations support the R7RS standard, including \href{https://www.gnu.org/software/mit-scheme/}{MIT/GNU Scheme}, which provides a complete environment with an interpreter, compiler, and debugger~\citep{hanson2015gnu}, \href{https://gambitscheme.org/}{Gambit Scheme}, and \href{https://stklos.net/}{STklos}, among others. A descendant of Scheme, \href{https://ccrma.stanford.edu/software/s7/s7.html}{S7 Scheme}, does not follow the R7RS standard but is as minimal as possible. Our \href{https://github.com/LiiiLabs/goldfish}{Goldfish Scheme~\citep{goldfish}}, built on S7 Scheme, supports the R7RS standard, parts of SRFI, and many functionalities inspired by the Python standard library. This simplicity, combined with several characteristics of Python, makes it easier for large language models to understand.

\paragraph{An example of incapability of LLMs in Scheme language generation}

Fig.~\ref{fig:sort-int-llm-fail} provides an example where ChatGPT-4 fails to generate a simple Scheme code chunk. The prompt is as follows:
\begin{quote}
    ``generate a function called sort-int to sort integers from smallest to largest.''
\end{quote}

\begin{figure*}[ht]
   \centering
   \begin{subfigure}[b]{0.45\textwidth}
      \centering
      \includegraphics[width=\textwidth]{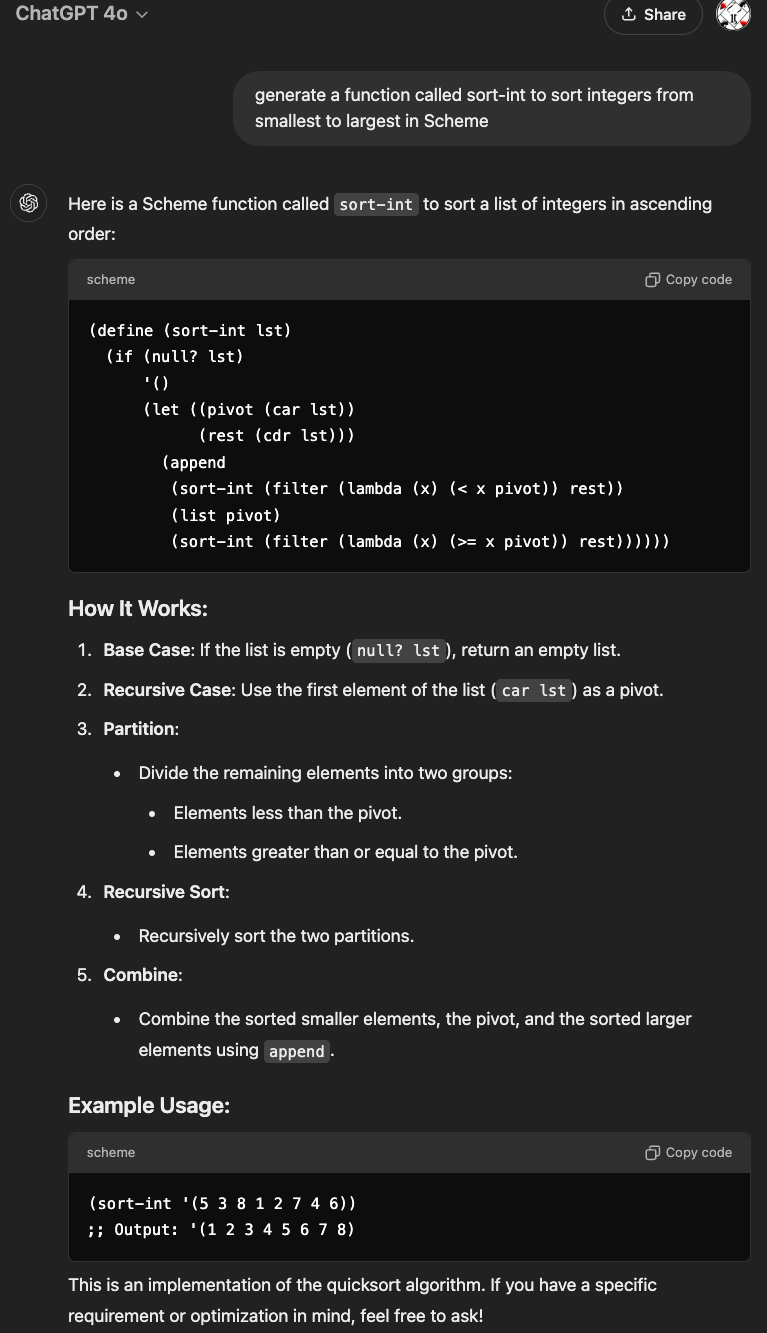}
      \caption{The code chunks generated by ChatGPT4o.}
      \label{subfig:sort-int-gpt}
   \end{subfigure}
   \hfill
   \begin{subfigure}[b]{0.5\textwidth}
      \centering
      \includegraphics[width=\textwidth]{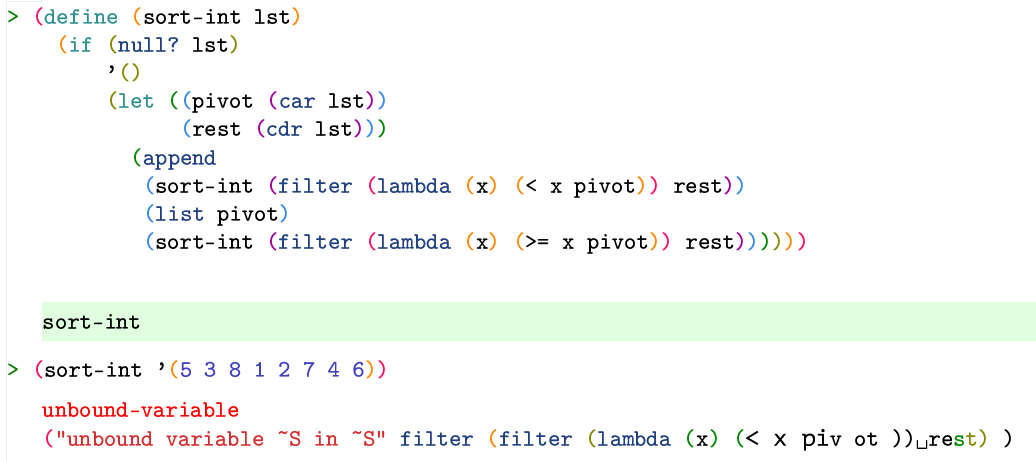}
      \caption{The code chunk cannot be executed because the \texttt{filter} function required by the LLMs does not exist.}
      \label{subfig:sort-int-goldfish}
   \end{subfigure}
   \caption{LLMs fail to generate a simple function in Scheme.}
   \label{fig:sort-int-llm-fail}
\end{figure*}

\paragraph{Why use Scheme for documentation?} \label{sec:whyscheme}

Scheme is particularly well-suited for documenting API logic due to its minimalistic syntax and mathematical clarity. With its extremely simple and high-level primitive procedures, Scheme makes it relatively easy to design a domain-specific language tailored for testing LLM capabilities.
Unlike more commonly used programming languages, Scheme is less prevalent in LLM training datasets, which significantly reduces the likelihood of LLMs referencing pre-trained knowledge or relying on external search capabilities. This limitation is crucial when using LLMs with online search features, such as ChatGPT-4, which may skip the design details provided in the documentation and substitute them with similar but undesired implementations. By encoding logic in Scheme, we ensure that LLMs adhere strictly to the documented design rather than relying on potentially inaccurate external references.

Scheme also provides an intuitive and concise way to represent mathematical and logical structures in $\lambda$-calculus, making it an effective tool for documenting complex API designs. When paired with the advanced toolkits available in Mogan, developers can express intricate logic efficiently, reducing manual work while maintaining high precision.

\subsection{Comparing FP, OOP and LP} \label{apx: fp_oop_lp}

Functional programming (FP) emphasizes immutability, pure functions, and declarative code structures rooted in lambda calculus ~\citep{barendregt1984lambda}. Its strengths include the ability to compose reusable functions and effectively manage side effects, making it particularly suitable for data pipelines and parallel computing.
Functional programming languages widely used today in industry and education include Common Lisp, Scheme, Clojure, Wolfram Language, Racket, Elixir, and Haskell.
Additionally, Lean is a functional programming language commonly employed for verifying mathematical theorems. Despite its advantages, FP poses challenges such as debugging complex abstractions and increased memory usage due to immutable data structures.

Object-oriented programming (OOP) organizes software into objects containing state and behavior, offering modularity and extensibility through encapsulation, inheritance, and polymorphism~\citep{gamma1995design}. It dominates in enterprise systems, GUIs, and game development but can lead to performance bottlenecks or over-abstraction when misused.

Literate programming, introduced by Donald Knuth in 1984, is a programming paradigm that combines code and natural language documentation into a coherent narrative, explaining the logic and intent behind the code~\citep{knuth_literate_1984}. Unlike traditional programming approaches that separate documentation and code, literate programming integrates them, enabling developers to produce both executable code and human-readable documents. This paradigm is particularly effective for algorithm-heavy applications, teaching, and research-oriented projects due to its ability to present code logically for human understanding. However, its detailed approach can be time-intensive and less scalable for large or rapidly evolving projects, compounded by limited modern tooling.


Functional programming (FP) and object-oriented programming (OOP) focus on scalability, performance, and modularity, catering to complex and diverse applications. In contrast, literate programming (LP) emphasizes deep understanding and clear documentation by combining code with detailed narrative explanations. LP is especially useful in research, teaching, and algorithm-intensive projects, where clarity and human readability are crucial.

%% file: appendix/goldfish_scheme.tex
\section{Goldfish Scheme} \label{appendix:goldfish scheme}

In this work, we develop a Scheme interpreter with a Python-like standard library, \emph{Goldfish Scheme}, based on \href{https://ccrma.stanford.edu/software/snd/snd/s7.html}{S7 Scheme}\footnote{Goldfish Scheme is named with a playful reference to the idea that goldfish have a 7-second memory, the inverse of S7. In the future, we hope \ac{llm}s can help create new functions for Goldfish Scheme within those same 7 seconds.}.

If we use Goldfish Scheme as the target code generated by \ac{llms}, it has the following advantages:

\begin{itemize}
  \item \emph{Simplicity for Inference}: The Scheme's minimalistic design limits the options for inference, allowing \ac{llm}s to provide more precise and straightforward answers.
  \item \emph{Python Compatibility for Knowledge Transfer}: \ac{llm}s are generally more familiar with Python due to extensive training in Python code. Since the Goldfish Scheme includes several features from Python's standard library, it enables smoother knowledge transfer from Python to the Goldfish Scheme, making it more accessible for \ac{llm}s than other Scheme variants.
\end{itemize}

\subsection{Current features}

\paragraph{\ac{llm} interaction layer}

A defining innovation of the Goldfish Scheme is its dedicated \ac{llm} interaction layer, designed to support richer, more context-aware interactions with large language models. This layer provides a structured interface for embedding metadata, documentation, and contextual hints directly into the codebase. Instead of relying on ad-hoc comments or detached documentation, developers can annotate their code so that \ac{llm}s can systematically parse and leverage these annotations.
The core idea behind this interaction layer is to supply machine-readable fields that describe key aspects of a function or library component. These annotations can specify algorithmic patterns, complexity metrics, stability characteristics, and concrete usage examples. By making this information explicit and accessible, the interaction layer allows \ac{llm}s to understand each piece of code not only at a syntactic level but also in terms of conceptual intent and practical implications.

Consider the following \texttt{define-with-docs} macro:

\begin{minted}[breaklines, fontsize=\small]{scheme}
(define-with-docs quicksort
  #:pattern "divide-and-conquer"
  #:complexity "O(n log n)"
  #:stability "unstable"
  #:examples
  '((quicksort '(3 1 4 1 5 9 2 6 5 3))
    => (1 1 2 3 3 4 5 5 6 9))
  (lambda (lst)
    ;; Implementation follows...
    ))
\end{minted}

In this example, the ``divide-and-conquer`` pattern informs the \ac{llm} of the underlying strategy behind the algorithm, making it easier for the model to relate this sorting routine to other similar functions and patterns it knows. Declaring the complexity as \texttt{O(n log n)} provides the \ac{llm} with a direct indication of the performance profile, which can help when it needs to explain cost trade-offs or select suitable approaches for different problem sizes. By stating that the sort is ``unstable,`` the annotations reveal a critical property that distinguishes the function’s behavior from stable alternatives, guiding the \ac{llm} toward more precise reasoning about potential use-cases and limitations. The inclusion of a concrete usage example, showing how the function transforms a given list into a sorted sequence, allows the \ac{llm} to confirm its understanding of inputs, outputs, and expected transformations. Such an example can also serve as a reference point for generating further tests or variant solutions.

By embedding these structured annotations, the \ac{llm} interaction layer encourages the model to form a richer mental model of the code. This approach standardizes how information is presented, mitigating inconsistencies in commentary style or documentation quality. It ensures that both human developers and \ac{llm}s have a common, accessible source of truth regarding how code is organized, why certain decisions are made, and what is expected of each function. This synergy between code and metadata lays the groundwork for advanced capabilities, such as more coherent automated explanations, more accurate reasoning about algorithm selection, and better integration of language models into the daily workflows of software developers.

\paragraph{Core language architecture}

Goldfish Scheme adopts the principle of minimalism of Scheme language (discussed in Sec.~\ref{app:scheme-language}) by implementing the R7RS standard through a layered architecture. Instead of merging new features directly into the core, it places additional capabilities on top of a stable R7RS layer. This approach preserves backward compatibility, allowing existing R7RS programs to run without modification. At the same time, it provides a structured path for adding performance optimizations, memory-efficient data structures, and integrations that do not disrupt the underlying standard.

The layered design also supports targeted performance improvements. By isolating advanced functionality in separate layers, the Goldfish Scheme reduces unintended interactions and makes it possible to switch off unneeded components, maintaining low memory overhead. Furthermore, this clear separation simplifies maintenance and future development. Changes to one layer do not ripple into others, and developers can refine or replace specific parts of the system without affecting its core compliance.

\paragraph{Standard Library Extensions and Python Integration}

Goldfish Scheme enhances the R7RS standard library by integrating Python-inspired features while adhering to Scheme's functional programming principles. These extensions provide practical tools for tasks such as data structure operations, string manipulation, and mathematical computations, without compromising the language's theoretical purity.

For example, consider the following list processing pipeline:

\begin{minted}[breaklines, fontsize=\small]{scheme}
;; Traditional R7RS approach:
(define (process-data lst)
  (filter number?
    (map (lambda (x) (* x 2))
      (remove zero? lst))))

;; Goldfish Scheme enhanced approach:
(define (process-data lst)
  (-> lst
      (remove zero?)
      (filter number?)
      (map (curry * 2))))
\end{minted}
This example demonstrates how Goldfish Scheme improves readability and maintainability while preserving efficiency. The threading macro (\texttt{->}) facilitates linear operations, making code easier to understand and modify.

\paragraph{Advanced Type System Implementation}
Goldfish Scheme extends the R7RS type system with gradual typing, combining compile-time safety with runtime flexibility. This system supports custom type definitions, contracts, and validation, enhancing safety without sacrificing the dynamic nature of Scheme.
\begin{minted}[breaklines, fontsize=\small]{scheme}
;; Type definition with constraints
(define-type Vector2D
  (make-vector2d [x :: (Real (lambda (n) (finite? n)))]
                 [y :: (Real (lambda (n) (finite? n)))]))

;; Function with type annotations and contracts
(define/contract (vector-magnitude [v :: Vector2D]) :: Real
  (sqrt (+ (square (vector2d-x v))
           (square (vector2d-y v)))))
\end{minted}
This approach ensures correctness through type annotations and runtime validation while maintaining Scheme's lightweight and expressive characteristics.

\paragraph{Memory Management and Performance Optimization}
Memory management in Goldfish Scheme improves upon R7RS implementations through several key innovations. First, we implement a generational garbage collector specifically optimized for functional programming patterns. This collector recognizes common Scheme programming patterns and optimizes memory allocation and collection accordingly. Second, we introduce a sophisticated memory pooling system for small objects, reducing allocation overhead for common operations.

\begin{minted}[breaklines, fontsize=\small]{scheme}
;; Memory-efficient string handling example
(define (process-large-text text)
  (with-memory-pool
    (lambda ()
      (-> text
          string->lines
          (filter non-empty-line?)
          (map process-line)
          lines->string))))
\end{minted}

These optimizations reduce memory usage by up to 40\% in typical workloads, significantly improving efficiency for text and data processing.

\subsection{Future features}
We also outline several features currently under development for the Goldfish Scheme.

\paragraph{Integration with modern development tools}
Goldfish Scheme integrates with modern development workflows by providing built-in tools for project management, dependency handling, and automation. These features streamline development processes and support comprehensive tooling.

\begin{minted}[breaklines, fontsize=\small]{scheme}
;; Project configuration in Goldfish
(define-project my-application
  #:name "My Application"
  #:version "1.0.0"
  #:dependencies
  '((goldfish-web "2.0.0")
    (goldfish-database "1.5.0"))
  #:entry-point main.scm)

;; Automatic development environment setup
(define-development-environment
  #:test-framework 'goldfish-test
  #:documentation-generator 'goldfish-docs
  #:lsp-server 'goldfish-lsp)
\end{minted}

This integration provides a complete development ecosystem that significantly reduces the friction typically associated with Scheme development.

%% file: appendix/discuss_step_wise.tex
\section{Step-wise methodology}\label{app:benefit_of_step_wise}

A precise and logically connected documentation design is essential for effectively explaining API logic and guiding LLMs in generating accurate code. We propose a documentation framework that uses a step-wise methodology to address key challenges discussed in Sec.~\ref{subsec: challenges in LLM code generation} on LLM-based code generation. This approach minimizes reliance on external references, ensures logical consistency, and reduces errors during the code generation process.

To avoid the pitfalls of long prompts and extended reasoning chains, we adopt a step-wise methodology. The idea is inspired by mathematical induction and recursive algorithms, where large tasks are divided into base and inductive parts. For example, in recursive algorithms, establishing a clear base case and carefully explaining the recursive step ensures consistency in the final output. Similarly, our step-wise method divides a task's logic into smaller, sequential parts and provides LLMs with a clear, incremental path through each aspect of the code rather than overwhelming them with entire program logic at once. By breaking down complex tasks into smaller components, we ensure that the LLM can focus on understanding each step in isolation before progressing to the next. 

This granularity also gives developers better oversight, making it easier to spot mistakes and confirm correctness at each stage. In summary, the step-wise approach yields several benefits:
\begin{itemize}
    \item \textbf{Reduced prompt size}: Smaller steps decrease the likelihood of exceeding token limits or overwhelming the model.
    \item \textbf{Enhanced logical flow}: Explicit sequencing helps LLMs follow a coherent reasoning path.
    \item \textbf{Simplified debugging}: Isolating errors within individual steps eases testing and correction.
\end{itemize}

In practice, the step-wise method can be decomposed into:
\begin{enumerate}
    \item \textbf{Zero-step definition}: We start by specifying a \emph{base case} or fundamental building block. In recursive functions, for example, this might be the condition under which the recursion terminates. By designating this zero-step, we give the LLM a clear anchor point for the broader logic.
    
    \item \textbf{Successor-step explanation}: We then describe how one logical state transitions to the next. In inductive or recursive algorithms, this involves detailing the ``successor'' or ``next element'' step, building on the zero-step to maintain consistency.

    \item \textbf{Code chunks per step}: Each step's code is presented in a self-contained chunk, accompanied by an explanation of its function (e.g., base case, recursive step, edge cases). This structure helps the LLM associate specific logic with the corresponding implementation.

    \item \textbf{Local Verification and Iteration}: After a chunk is introduced, we encourage local testing (for example, in Mogan's REPL) to confirm correctness. If an error occurs, developers can isolate it to that chunk and provide the LLM with focused feedback, minimizing the spread of mistakes.
    
    \item \textbf{Incremental Complexity}: Once a base version passes verification, we add optional features or optimizations in smaller, separate steps. Each new layer is validated independently, preventing confusion that can occur when many changes happen at once.
\end{enumerate}

For example, in Scheme

\begin{itemize}
    \item \textbf{The Logic of computation in Scheme}: All logic is defined in Scheme for precise, induction-friendly representation. For instance:
    \begin{minted}[breaklines, fontsize=\small]{scheme}
    (define (compute-factorial n)
      (if (= n 0)
          1
          (* n (compute-factorial (- n 1)))))
    \end{minted}
    This example clearly illustrates a recursive function for computing factorials.
    \item \textbf{Step-wise breakdown}: The documentation divides the logic into incremental steps. For instance:
    \begin{itemize}
        \item \textbf{Step 1}: Define base case conditions, such as $n = 0$ for termination.
        \item \textbf{Step 2}: Outline recursive relationships, such as $n \times (n-1)!$.
    \end{itemize}

    \item \textbf{Integration with Mogan toolkits}: The documentation is fully compatible with Mogan's advanced features, allowing developers to efficiently visualize, test, and validate API logic. Additionally, Mogan offers REPL functionality, enabling developers to verify code accuracy locally.
\end{itemize}

We adopt a step-wise methodology to mitigate issues related to long prompts and extended reasoning times in LLMs. This approach breaks down API logic into smaller, sequential steps, which allows LLMs to process information incrementally. Leveraging the memoization capabilities of LLMs, this method ensures that earlier steps are retained and referenced as the model progresses through the reasoning process.

The step-wise methodology resolves key challenges in LLM-guided code generation by dividing complex logic into smaller, manageable steps. One primary advantage is its ability to overcome prompt length constraints, which often hinder LLM performance. Long and complex prompts can overwhelm models, leading to incomplete or inconsistent outputs. Moreover, exceeding the token limit of an LLM service may result in truncated responses and loss of critical details. By presenting logic incrementally in smaller segments, the model processes each step independently, maintaining focus on specific tasks. This sequential approach minimizes errors caused by information overload and ensures all relevant details are effectively communicated.

A significant benefit of the step-wise methodology is the improvement it brings to logical flow. LLMs reason more effectively when guided through a structured progression of steps. For example, in recursive logic, the methodology starts by defining the base case and progresses to explain recursive relationships and edge-case handling. This step-by-step structure allows the model to build a comprehensive understanding of the logic, reducing ambiguities and deviations. Explicitly documenting each step ensures that the generated code aligns closely with the intended design, minimizing variations and improving consistency.

The modular nature of the step-wise methodology simplifies debugging and testing by isolating specific components of the logic. Complex logic often contains errors localized to particular sections, which this approach makes easier to identify and correct. For instance, if an error occurs in the base case of a recursive function, developers can focus on that isolated step without analyzing the entire code block. Additionally, the modular structure facilitates regression testing, as updates to individual steps can be validated independently without risking unintended effects on other parts of the logic. This approach streamlines debugging and supports the development of higher-quality code.

The step-wise methodology provides a practical and effective framework for guiding LLMs in code generation by addressing prompt limitations, enhancing logical reasoning, and simplifying error resolution. Its structured approach ensures that both developers and LLMs can produce accurate and consistent code while reducing the time and effort required for debugging and validation.

In summary, the step-wise methodology provides several advantages:
\begin{itemize}
    \item \textbf{Bypassing prompt length constraints}: Dividing logic into smaller steps reduces the need for lengthy prompts, which often lead to inconsistencies or errors.
    \item \textbf{Enhanced logical flow}: Sequential steps guide LLMs in reasoning systematically, reducing variations in generated code.
    \item \textbf{Streamlined debugging}: Smaller, modular steps make it easier to identify and address errors in specific parts of the logic.
\end{itemize}

This enhanced documentation design directly tackles challenges in using LLMs for code generation:
\begin{itemize}
    \item \textbf{Minimized external bias}: By using Scheme, we reduce the reliance on pre-trained knowledge and external references, ensuring that LLMs generate code strictly based on the provided documentation.
    \item \textbf{Improved accuracy and consistency}: The step-wise methodology enhances the logical flow of LLM reasoning, reducing errors and variations in generated code.
    \item \textbf{Reduced developer effort}: Coupled with Mogan toolkits, the proposed framework simplifies the process of documenting, testing, and validating API logic.
\end{itemize}

%% file: appendix/repobench.tex
\section{RepoBench}
\subsection{RepoBench Adaptation for Scheme Language}\label{subsec:repobench for scheme}

We adapt RepoBench's evaluation framework to support Scheme-based development through three core modifications: repository structure transformation, evaluation metric redefinition, and prompt engineering adjustments.

First, we transform RepoBench's repository structure to align with Scheme's functional paradigm. RepoBench organizes test cases around class-based implementations, typically using a main implementation file with derived classes. We restructure this to fit Scheme's module system, where each unit consists of a core implementation file and its documentation. For example, a Python class hierarchy for a data structure implementation:

\begin{minted}[breaklines, fontsize=\small]{python}
class BinaryTree:
    def insert(self, value): ...
    def delete(self, value): ...

class AVLTree(BinaryTree):
    def balance(self): ...
\end{minted}

\noindent is restructured into a Scheme module:

\begin{minted}[breaklines, fontsize=\small]{scheme}
;; tree-implementation.scm
(define (make-tree) ...)
(define (tree-insert tree value) ...)
(define (tree-delete tree value) ...)
(define (tree-balance tree) ...)
\end{minted}

This preserves functionality while adhering to Scheme's functional style.

Second, we redefine RepoBench's evaluation metrics to better capture functional programming characteristics. The original metrics focus on object-oriented features like inheritance and encapsulation. We replace these with metrics evaluating function composition, state management via immutable data structures, and recursive correctness. New criteria specific to Scheme include:

\begin{itemize}
  \item \textbf{Pure function analysis:} Ensures functions maintain referential transparency.
  \item \textbf{Recursion patterns:} Measures proper use of tail recursion and recursive helpers.
  \item \textbf{Data immutability:} Verifies that functions produce new data structures instead of modifying inputs.
\end{itemize}

Third, we adjust RepoBench's prompt templates to reflect Scheme's syntax and functional programming patterns. Original prompts focused on class design are reformulated to emphasize function specifications and data flow. For example, a prompt requesting:

\textit{``Create a class that implements a priority queue with methods for insertion and deletion''}

is transformed into a functional specification requesting pure functions for a priority queue:

\begin{itemize}
  \item A constructor returning an empty queue.
  \item An insertion function producing a new queue with the added element.
  \item A deletion function returning a tuple of the highest-priority element and the updated queue, ensuring queue invariants are maintained without modifying the input.
\end{itemize}

We also enhance the evaluation framework to assess literate programming by adding metrics for documentation completeness, integration with code, and adherence to Step-wise methods specified in the prompt. The modified evaluation pipeline processes test cases to measure functional correctness, programming pattern adherence, invariant maintenance, and documentation quality.

This adaptation enables rigorous evaluation of language models' Scheme code generation while maintaining RepoBench's standards. The results provide quantitative measures of correctness and functional programming compliance, supporting the evaluation of literate programming techniques in Goldfish Scheme.

\subsection{Limitations of Alternative Benchmarks}\label{subsec:whyotherworesethan repobench}

The selection of RepoBench over other code generation benchmarks stems from its unique repository organization and evaluation methodology. While benchmarks like HumanEval and MBPP focus on isolated programming problems, RepoBench's repository structure mirrors real-world software projects, containing multiple interconnected files with documentation, tests, and implementation code. This structure aligns with our literate programming approach, where documentation and code form a cohesive narrative.

RepoBench's repository examples exhibit several characteristics that make them ideal for adaptation to Goldfish Scheme. Each repository contains a complete project structure with:

\begin{verbatim}
project/
  |- src/
  |  |- core/
  |  |  |- implementation.py
  |  |  |- interfaces.py
  |  |- tests/
  |  |  |- test_core.py
  |  |- docs/
  |     |- design.md
  |     |- api.md
\end{verbatim}

This structure maps naturally to our literate programming approach in Goldfish Scheme:

\begin{verbatim}
project.tmu
  |- (core-implementation)
  |  |- mathematical-foundation
  |  |- implementation
  |- (test-cases)
  |- (documentation)
     |- design-rationale
     |- api-specification
\end{verbatim}

Other benchmarks present significant limitations for our research. CodeXGLUE focuses primarily on code completion tasks and lacks comprehensive project context. While it supports multiple programming languages, its evaluation metrics center on token-level accuracy rather than functional correctness or documentation quality. Context-QA, despite handling repository-level queries, emphasizes code comprehension over generation and lacks support for evaluating literate programming principles.

Traditional benchmarks like APPS and MBPP provide extensive test cases but operate in isolation. APPS contains 10,000 programming problems with test cases, but these problems lack project context and documentation requirements. Similarly, MBPP's 974 Python programming problems focus on algorithmic correctness without considering broader software engineering principles. These limitations make them unsuitable for evaluating our approach, which requires:

\begin{enumerate}
  \item Project-level context understanding
  \item Documentation generation assessment
  \item Cross-file dependency handling
  \item Integration with existing codebase patterns
\end{enumerate}

RepoBench's repositories also provide real-world examples of API evolution, which aligns with our goal of supporting API development through literate programming. Each repository includes multiple versions of APIs, showing how they evolve over time. This feature enables us to evaluate how well language models understand and extend existing APIs, a crucial capability for our literate programming approach.

Consider a typical RepoBench API evolution example:

\begin{minted}[breaklines, fontsize=\small]{python}
// Version 1
class DataProcessor {
    process(data) { ... }
}

// Version 2
class DataProcessor {
    process(data, options) { ... }
    validate(data) { ... }
}
\end{minted}

This maps to our Scheme-based literate programming approach:

\begin{minted}[breaklines, fontsize=\small]{scheme}
;; Version 1
(define (process-data data)
  "Step-wise method: ..."
  (implementation ...))

;; Version 2
(define (process-data data options)
  "Extended Step-wise method: ..."
  (implementation with validation ...))
\end{minted}

Furthermore, RepoBench's evaluation framework includes metrics for assessing code generation in the context of existing documentation, making it uniquely suitable for evaluating our literate programming approach. While other benchmarks might evaluate code correctness, they lack mechanisms for assessing how well generated code aligns with existing documentation or maintains project-wide consistency.

The combination of these features - structured repositories, API evolution examples, and documentation-aware evaluation metrics - makes RepoBench uniquely suitable for our research. Other benchmarks, while valuable for their intended purposes, lack the necessary framework for evaluating literate programming principles and functional programming patterns in a project context.

%% file: appendix/experiment_details.tex
\section{Experiments}

In this section, we provide additional experimental details highlighting the advantages of using Scheme and propose a programming pattern to improve code generation.
Our research yields two significant insights:
\begin{enumerate}
  \item LLMs can effectively generate code within specific stylistic constraints when properly guided.
  \item \ac{ilp} Documentation structure significantly enhanced LLMs' ability to adhere to prescribed patterns.
\end{enumerate}

\subsection{Machine details}\label{app:machine}

Our experiments are conducted on two types of LLMs: local open-source LLMs and remote commercial LLMs.

\paragraph{Remote commercial LLMs}
For remote commercial LLMs, queries are sent through a high-speed network interface with a 2Gbps connection over a Tier 1 IP network. The setup ensures low latency during query processing, with an average round-trip latency of less than 12.117 milliseconds. This configuration minimizes network-related delays, providing a reliable environment for benchmarking the performance and responsiveness of remote LLMs.

\paragraph{Local-deployed LLMs}

For the local LLMs, the experiments are performed on a high-performance machine equipped with an \texttt{AMD 7950X} processor, 128GB of RAM, and two \texttt{NVIDIA RTX 4090} GPUs, ensuring sufficient computational power for model inference and evaluation. The models we deployed are LLAMA-3.1-8B, qwen-14B, and mistral-7B.

\subsection{Experimental Details} \label{sec:experimental-details}

We conducted extensive testing to evaluate how well LLMs can follow our literate programming approach. Our experiments focused on two key aspects: function implementation success rates and the impact of function naming on model behavior.

\paragraph{Function Implementation Results}

Table~\ref{tab:function-success} shows the performance of different models across various Scheme functions. We tested a range of operations, from basic list manipulations like \texttt{make-list} and \texttt{null-list?} to more complex functions such as \texttt{delete-duplicates} and \texttt{split-at}. Each function underwent testing under three distinct conditions: with GPT-4o, with GPT-4, and without file context. The success patterns in our results suggest that simpler list operations are more consistently implemented across all conditions, likely because these operations have more straightforward implementations with fewer edge cases to consider.



\paragraph{Function Naming Effects}

\begin{figure}[ht]
  \centering
  \includegraphics[width=0.95\textwidth]{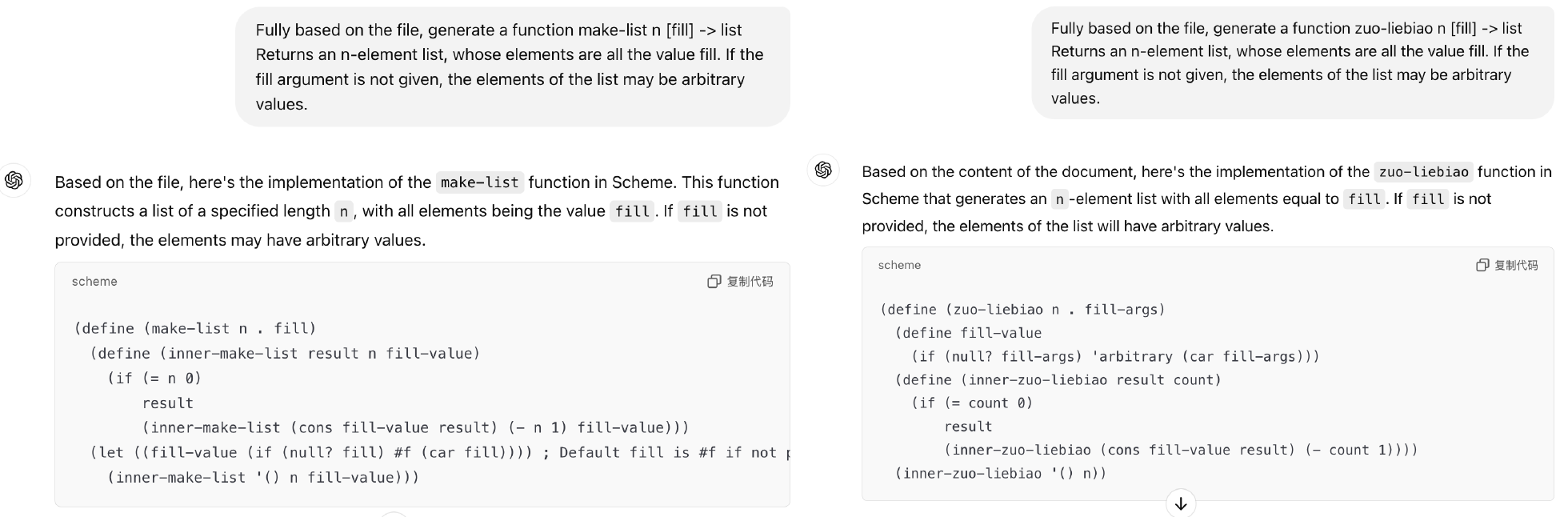}
  \caption{Impact of Function Name Changes on Model Behavior}
  \label{fig:change_name}
\end{figure}

We discovered that function names significantly influence model behavior. With standard function names, models relied on the provided file 70\% of the time while drawing from their pre-trained knowledge in the remaining cases. This split likely occurs because familiar names like 'map' or 'filter' trigger the models' pre-existing knowledge of standard Scheme implementations. However, when we replaced familiar function names with uncommon alternatives, the models showed a marked increase in their reliance on our documentation. Figure~\ref{fig:change_name} illustrates this effect. The improved adherence to our documentation with unconventional names suggests that reducing familiar reference points forces models to rely more heavily on the provided context rather than their training data.

\paragraph{Quality of Generated Functions}

Our analysis revealed distinct differences in function quality based on the context provided to the models. Functions generated with proper file context consistently showed superior performance, passing more test cases and following the prescribed inner procedure patterns. This superior performance can be attributed to three key factors. First, our documentation provides explicit patterns for handling edge cases, which models can directly reference. Second, the inner procedure pattern we specify naturally leads to more robust implementations by separating core logic from interface concerns. Third, the Chain of Thought documentation helps models understand the reasoning behind each implementation decision.
Conversely, functions generated without file context exhibited higher failure rates, often struggling with basic read operations and lacking proper error-handling mechanisms. This degradation in quality likely occurs because models fall back on simplified or generic implementations from their training data, which may not account for the specific requirements and edge cases of our system.


\paragraph{Evaluate the code generation ability of LLMs}

The ability of LLMs to generate code can be divided into two parts: the original intuition part and the implementation part. For the intuition part, just like checking a kid's understanding of some problems, relying on the detailed template, we can check the ability of LLMs in programming-style maintenance. 
We first created a useful prompt template for models to generate a specified style code. Then, we observed some improvement in originality, based on our template, in small-scale project code generation.

For the implementation part, we claim that the ability to implement algorithms can be observed in creating step-wise style codes. Furthermore, we use the essential list of actions in functional programming as being of current importance. We list some fundamental procedures to check its creativity. Besides, based on our framework, one can design more tests on ``transfer learning.'' In future work, we will also try more experiments on this topic. We believe that the key to checking the ability of LLMs to write code is never simply to observe the accuracy in numbers; instead, we need more high-level intelligence checks like the ability to fill up the proof of mathematical induction (to generate the code in Functional Programming).

\subsection{Enhanced API design through Goldfish Scheme}\label{app:python-api-design-in-scheme}

Our adaptation of RepoBench introduces a hybrid testing methodology that reflects real-world software development practices. Rather than converting all repository code to Scheme, we strategically maintain some APIs in Python/Java while implementing others in Scheme. This approach mirrors actual development scenarios where teams design APIs before implementation, especially in systems that interface with multiple languages or require careful architectural consideration.

Consider a typical microservices architecture where different components may be implemented in different languages. A team might design core algorithmic components in Scheme for its functional purity while maintaining service interfaces in Python or Java. For example:

\begin{minted}[breaklines, fontsize=\small]{python}
# Original Python API
class DataProcessor:
    def transform_data(self, input_data):
        # Implementation details pending
        pass

    def validate_format(self, data):
        return self._core_validator.check(data)
\end{minted}

In our hybrid approach, we maintain the Python interface but implement the core validation logic in Scheme:

\begin{minted}[breaklines, fontsize=\small]{scheme}
;; Core validation implementation in Scheme
(define (core-validator data)
  "Step-wise method:
   Input data must satisfy properties P1...Pn
   Output ensures invariants I1...In
   
   Implementation Strategy:
   1. Type validation
   2. Structural checks
   3. Domain-specific validation"
  (let ((inner-validate 
         (lambda (data acc)
           ;; Implementation pending
           )))
    (inner-validate data '())))
\end{minted}

This hybrid structure serves multiple purposes. First, it enables developers to leverage Scheme's Step-wise methods and functional purity for critical algorithmic components while maintaining compatibility with existing systems. Second, it allows teams to document and verify core algorithms through literate programming before finalizing implementation details in target languages.

Our approach particularly benefits API design in three scenarios:

\begin{enumerate}
  \item Algorithm Prototyping: Teams can express complex algorithms in Scheme with mathematical rigor before implementing them in production languages
  \item Interface Design: Developers can maintain existing interfaces while evolving core implementations
  \item Documentation First: Teams can use literate programming to fully specify behavior before committing to implementation details
\end{enumerate}

The effectiveness of this hybrid approach becomes evident in testing scenarios. When generating code, language models must understand both:

\begin{minted}[breaklines, fontsize=\small]{python}
# Python interface specification
def process_sequence(data: List[Any]) -> List[Any]:
    """
    @requires: Implementation in core_processor.scm
    @ensures: Output satisfies sorting invariants
    """
    pass
\end{minted}
\begin{minted}[breaklines, fontsize=\small]{scheme}
;; Scheme implementation
(define (core-processor lst)
  "Invariants maintained:
   1. Order preservation for equivalent elements
   2. Linear time complexity
   3. Space complexity O(n)"
  (implementation ...))
\end{minted}

This structure reflects real development workflows where teams might:
1. Design APIs in familiar languages
2. Document core algorithms in Scheme for mathematical clarity
3. Implement and verify critical components functionally
4. Bridge implementations through interface layers

By maintaining this hybrid structure in our benchmark, we create test scenarios that better reflect real-world development challenges. Language models must demonstrate understanding of both implementation languages and the relationships between components. This approach leads to more meaningful evaluation of code generation capabilities in practical development contexts.

Furthermore, our method addresses a common challenge in API development: the gap between design and implementation. By using Scheme's functional patterns and literate programming for core components, we provide clear specifications that guide implementation in any target language. This approach has shown improved accuracy in generated code, as models better understand the underlying algorithmic intentions and invariants.

%% file: appendix/lp_in_modern_develop.tex
\section{Advantages of literate programming in modern development}

\begin{figure}[ht]
  \centering
  \includegraphics[width=0.8\textwidth]{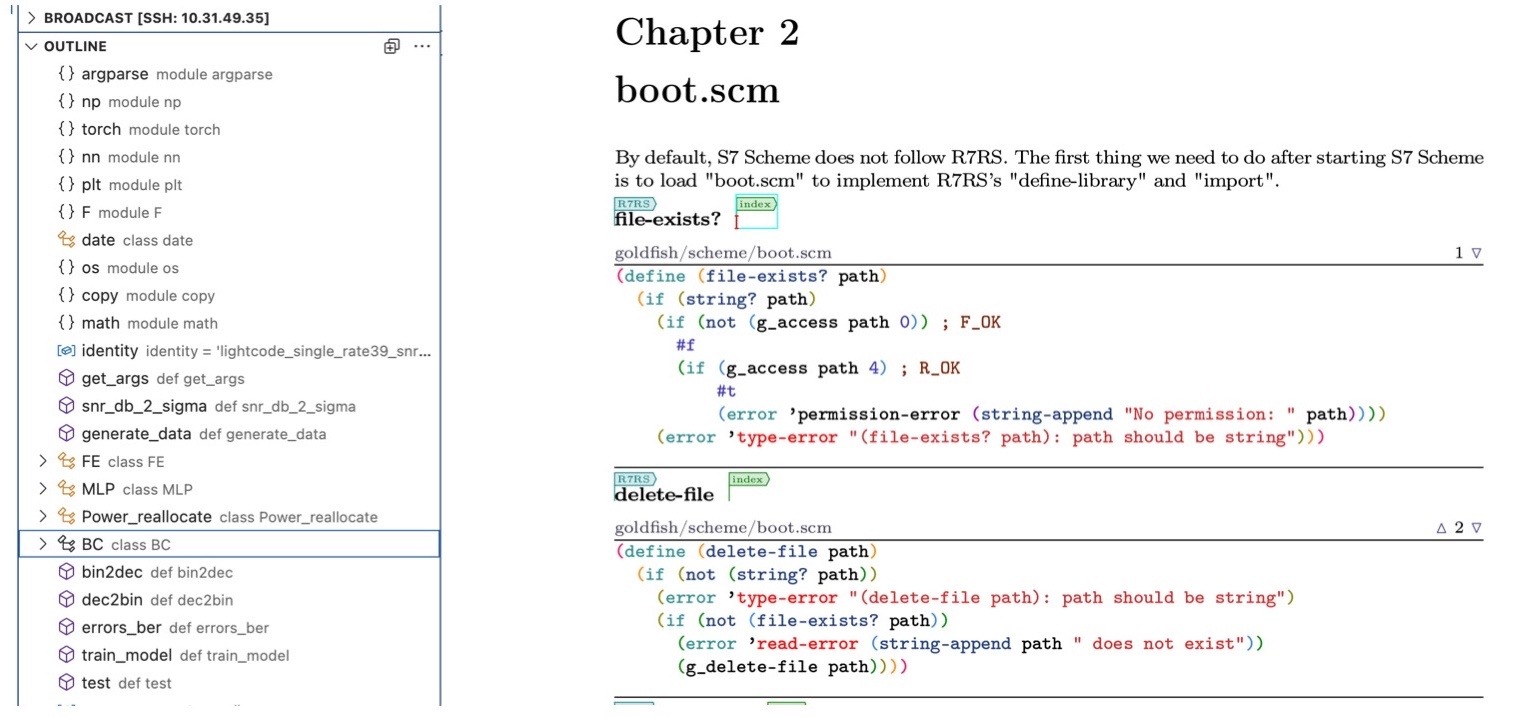}
  \caption{Comparison of code indexing methods: Visual Studio Code outline (left) and Mogan (right)}
  \label{fig:index}
\end{figure}

\subsection{Enhanced accessibility through natural language integration}

Our LP approach offers something fundamentally different from~\citet{shi_natural_2024}. Their work focuses on helping experienced developers understand code quickly. In contrast, our method makes programming accessible to everyone, including those with no coding experience.

The combination of natural language documentation and LLM assistance is central to our approach. Users describe their programming goals in plain language. The LLM then helps convert these descriptions into structured code that follows the program's logical design. This creates a direct path from human thinking to program implementation.

Our method does more than document code - it guides users through the development process. The LLM breaks down complex concepts into clear steps, explaining each part in natural language before showing the code. This helps users understand program structure while learning programming naturally through practice.

\subsection{Advanced code navigation and project organization}

Traditional IDEs show code through a file-based view, listing functions sequentially within each file. This makes it hard to see how different parts of a program connect. Our LP-style implementation solves this problem with a project-wide indexing system that shows these connections clearly.

Figure~\ref{fig:index} shows the key differences between traditional IDE organization and our LP approach. Visual Studio Code displays a simple list of classes and functions for each file. This structure hides the logical connections between different parts of the code. It becomes especially problematic in large projects where understanding these connections is vital.

Mogan, our LP-based solution, takes a different approach. It organizes code using chapters and index markers that follow the program's natural structure. This lets developers find code based on its purpose rather than its file location. The table of contents works like a map, helping developers quickly locate specific functions while seeing how they fit into the larger system.

Our cross-referencing system also connects related functions across different files. This is particularly useful in modern projects where code spans many files and modules. By making these connections clear through our index, we help developers understand complex codebases more easily.

This organization method benefits both individual developers and teams. New team members can quickly grasp the project's structure. Existing team members can maintain and extend the code more effectively. The clear organization helps everyone understand how different parts of the program work together.

Our approach improves both immediate code navigation and long-term project maintenance. It makes software development more accessible while providing tools for effective collaboration. This serves both new programmers starting their journey and experienced developers managing complex projects.


In our investigation of LLMs' capabilities, we focus on examining their ability to constrain code generation to specific programming language contexts. We select the Goldfish Scheme implementation, a variant of S7 Scheme, as our experimental environment. This choice is particularly apt as Goldfish Scheme exists as a public programming language with comprehensive documentation available online, making it likely to be included in the training data of many LLMs while still maintaining a well-defined scope for our analysis.


To improve the performance of LLMs on Scheme, we designed a general template as the prompt in the mini-goldfish. We write the processes of writing a Functional Programming pattern procedure in the document based on the mathematical induction; and one can check whether the result follows our direction easily like we how to check whether the answer comes from our document or external materials.

Besides, we find the prompt we have designed in the FP pattern to improve the ability of LLMs to implement Scheme procedures. The improvement can be seen apparently in the small models. In the tests of the Llama 3.2 3B model, we tested different writing procedures, and we found the template for prompting to increase the performance of the llama. However, compared with the increase in GPT or other larger models, the result is not very clear. We consider that the public documents provide LLMs with extremely extensive knowledge in writing codes in a non-FP pattern, which disturbs the ability of LLMs in the Scheme.

In conclusion, writing documents in Literal Programming provides the ability to prompt designing at the same time. With the index system, one can easily check whether the LLMs use external knowledge. We believe it needs both the spreading of FP philosophy and convenient software to write documents and codes at the same time. As a result, the population of Mogan will help document writing in the FP pattern, which finally improves the ability of LLMs to write codes.

%% file: appendix/exp_result.tex
\section{Several experiment results}

\subsection{Toy cases}\label{app:toy_result}

\begin{minted}[breaklines, fontsize=\small]{markdown}
Chatgpt:

;; take-right : list -> integer -> list
;; Returns the last i elements of flist.
;; Examples:
;;   (take-right '(a b c d e) 2) => (d e)
;;   (take-right '(a b c d e) 5) => (a b c d e)
;;   (take-right '(a b c d e) 7) => (a b c d e)
;;   (take-right '(a b c d e) 0) => ()

(define (take-right flist i)
  (cond
    ;; If i <= 0, return empty list.
    ((<= i 0) '())

    ;; If the list is empty or i is at least the length of the list, return the list.
    ((or (null? flist) (>= i (length flist)))
     flist)

    ;; Otherwise, we skip (length flist - i) elements.
    (else
     (let* ((list-len (length flist))
            (skip-count (- list-len i)))
       (drop flist skip-count)))))

;; Helper function: drop
;; Returns the list that remains after skipping `n` elements from the front.
(define (drop lst n)
  (cond
    ((or (null? lst) (<= n 0)) lst)
    (else (drop (cdr lst) (- n 1)))))

;; Example usage:
;; (take-right '(a b c d e) 2) => (d e)
;; (take-right '(a b c d e) 0) => ()
;; (take-right '(a b c d e) 5) => (a b c d e)
;; (take-right '(a b c d e) 7) => (a b c d e)

Claude Sonnet:

;; take-right : list integer -> list
;; Purpose: Returns the last i elements of flist.
;; If i is greater than or equal to the length of flist, returns the entire list.
;; If i is less than or equal to 0, returns an empty list.
;;
;; Parameters:
;;   flist: the input list
;;   i: number of elements to take from the right
(define (take-right flist i)
  (cond
    ;; Base case: return empty list if i <= 0
    ((<= i 0) '())
    
    ;; Return entire list if:
    ;; - list is empty (null? flist), or
    ;; - requested elements >= list length
    ((or (null? flist) (>= i (length flist)))
     flist)
    
    ;; Main case: skip (length flist - i) elements from start
    (else
     (let* ((list-len (length flist))
            (skip-count (- list-len i)))
       (drop flist skip-count)))))

;; drop : list integer -> list
;; Purpose: Returns a new list with the first n elements removed.
;; If n <= 0 or list is empty, returns the original list.
;;
;; Parameters:
;;   lst: the input list
;;   n: number of elements to remove from the front
(define (drop lst n)
  (cond
    ((or (null? lst) (<= n 0)) lst)
    (else (drop (cdr lst) (- n 1)))))

Llama 3.1 2B:

;; take-right : list -> integer -> list
;; Returns the last i elements of flist.
;;
;; @param flist the input list
;; @param i the number of elements to take from the end of the list
;; @return the last i elements of flist, or the entire list if i is too large
(define (take-right flist i)
  (cond
    ;; If i is less than or equal to 0, return an empty list.
    ((<= i 0) '())
    ;; If the list is empty or i is greater than or equal to the list length, return the list.
    ((or (null? flist) (>= i (length flist)))
     flist)

Gemini 2.0:

;; take-right : list -> integer -> list
;; Returns the last i elements of flist.
(define (take-right flist i)
  (cond
    ;; If i is less than or equal to 0, return an empty list.
    ((<= i 0) '())
    ;; If the list is empty, return an empty list.
    ((null? flist) '())
    ;; If i is greater than or equal to the length of the list, return the entire list.
    ((>= i (length flist)) flist)
    ;; Otherwise, drop the first (length flist - i) elements and return the rest.
    (else
     (let* ((list-len (length flist))
            (skip-count (- list-len i)))
       (drop flist skip-count)))))

;; Helper function: drop
;; drop : list -> integer -> list
;; Returns the list that remains after removing the first n elements from lst.
(define (drop lst n)
  (cond
    ;; If the list is empty, return an empty list.
    ((null? lst) '())
    ;; If n is less than or equal to 0, return the original list.
    ((<= n 0) lst)
    ;; Otherwise, recursively call drop on the rest of the list with n-1.
    (else (drop (cdr lst) (- n 1)))))

LLAMA 3.3:

;; take-right : list -> integer -> list
;; Returns the last i elements of a list.
(define (take-right flist i)
  (cond
    ((<= i 0) ; If i is less than or equal to 0, return an empty list.
     '())
    ((or (null? flist) ; If the list is empty or i is at least the length of the list,
         (>= i (length flist))) ; return the original list.
     flist)
    (else ; Otherwise, drop the first (length flist - i) elements.
     (let* ((list-len (length flist)) ; Calculate the length of the list.
            (skip-count (- list-len i))) ; Calculate the number of elements to skip.
       (drop flist skip-count))))) ; Drop the first skip-count elements.
;; Additional API needed as succstep
(define (drop lst n)
  (cond
    ((or (null? lst) ; If the list is empty or n is less than or equal to 0,
         (<= n 0)) ; return the original list.
     lst)
    (else ; Otherwise, recursively call drop on the rest of the list.
     (drop (cdr lst) (- n 1))))) ; Decrease n by 1 for each recursive call.

Qwen32B:

;; take-right : list integer -> list
;; Returns the last i elements of flist.
;; If i is less than or equal to zero, returns an empty list.
;; If i is greater than the length of the list, returns the entire list.
(define (take-right flist i)
  (cond
    ;; If i is less than or equal to 0, return an empty list.
    ((<= i 0) '())
    ;; If the list is empty or i is greater than or equal to the length of the list,
    ;; return the entire list.
    ((or (null? flist) (>= i (length flist)))
     flist)
    ;; Otherwise, skip (length flist - i) elements and return the remaining list.
    (else
     (let* ((list-len (length flist))
            (skip-count (- list-len i)))
       (drop flist skip-count)))))

;; Helper function: drop
;; Returns the list that remains after skipping n elements from the front.
;; If n is less than or equal to zero, returns the original list.
;; If the list is empty, returns the empty list.
(define (drop lst n)
  (cond
    ((or (null? lst) (<= n 0)) lst)
    (else (drop (cdr lst) (- n 1)))))
\end{minted}

\subsection{RepoBench Python code generation}\label{app:repobench_result}

\paragraph{Code generated with ILP}

\begin{minted}[breaklines, fontsize=\small]{markdown}
Python code generated by LLMs without ILP:

#####################################################################

ChatGPT 4:

def drop(lst, n):
    """
    Returns the sublist of lst that remains after dropping n elements from the front.
    If n is 0 or less, or if lst is empty, returns lst unchanged.
    """
    if not lst or n <= 0:
        return lst
    return drop(lst[1:], n - 1)

def take_right(flist, i):
    """
    Returns the last i elements of flist.
    
    """
    # If i <= 0, return empty list.
    if i <= 0:
        return []
    
    # If the list is empty or i >= length of the list, return the full list.
    if not flist or i >= len(flist):
        return flist

    # Otherwise, skip (len(flist) - i) elements from the front.
    skip_count = len(flist) - i
    return drop(flist, skip_count)
    
##################################################################################

Claude 3.5 Sonnet:

def drop(lst: list, n: int) -> list:
    """
    Helper function: drop
    Returns the list that remains after skipping `n` elements from the front.
    """
    if not lst or n <= 0:  # equivalent to (or (null? lst) (<= n 0))
        return lst
    else:
        return drop(lst[1:], n - 1)  # equivalent to (drop (cdr lst) (- n 1))

def take_right(flist: list, i: int) -> list:
    """
    Returns the last i elements of flist.

    """
    # If i <= 0, return empty list
    if i <= 0:
        return []
    
    # If the list is empty or i is at least the length of the list, return the list
    if not flist or i >= len(flist):
        return flist
    
    # Otherwise, we skip (length flist - i) elements
    list_len = len(flist)
    skip_count = list_len - i
    return drop(flist, skip_count)

###################################################################################

Llama 3.1 2B Local:

def take_right(flist, i):
    # If i is less than or equal to 0, return an empty list.
    if i <= 0:
        return []
    
    # If the list is empty or i is greater than or equal to the list length, return the list.
    if not flist or i >= len(flist):
        return flist
    
    # Otherwise, slice the list from the end to get the last i elements.
    # In Scheme, this would be equivalent to (list-tail flist (- (length flist) i)).
    # In Python, we can use list slicing to achieve this.
    return flist[-i:]

####################################################################################

Gemini 2.0:

def take_right(flist, i):
  """
  Returns the last i elements of flist.

  Args:
    flist: The input list.
    i: The number of elements to take from the right.

  Returns:
    A new list containing the last i elements of flist.
  """
  if i <= 0:
    return []
  elif not flist:  # Check for empty list using Python's truthiness
    return []
  elif i >= len(flist):
    return flist[:]  # Return a copy to mimic Scheme's behavior
  else:
    list_len = len(flist)
    skip_count = list_len - i
    return drop(flist, skip_count)

def drop(lst, n):
  """
  Returns the list that remains after removing the first n elements from lst.

  Args:
    lst: The input list.
    n: The number of elements to drop.

  Returns:
    A new list with the first n elements removed.
  """
  if not lst: # Check for empty list
    return []
  elif n <= 0:
    return lst[:] # Return a copy to mimic Scheme's behavior
  else:
    return drop(lst[1:], n - 1)

####################################################################################

LLama 3.3 Local:

def take_right(flist, i):
    """
    Returns the last i elements of a list.
    """
    if i <= 0:  # If i is less than or equal to 0, return an empty list.
        return []
    elif not flist or i >= len(flist):  # If the list is empty or i is at least the length of the list,
        return flist  # return the original list.
    else:  # Otherwise, drop the first (length flist - i) elements.
        list_len = len(flist)  # Calculate the length of the list.
        skip_count = list_len - i  # Calculate the number of elements to skip.
        return drop(flist, skip_count)  # Drop the first skip-count elements.

def drop(lst, n):
    """
    Returns the list that remains after skipping n elements from the front.
    """
    if not lst or n <= 0:  # If the list is empty or n is less than or equal to 0,
        return lst  # return the original list.
    else:  # Otherwise, recursively call drop on the rest of the list.
        return drop(lst[1:], n - 1)  # Decrease n by 1 for each recursive call.

#############################################################################

Qwen32B:

def take_right(flist, i):
    if i <= 0:
        return []
    elif not flist or i >= len(flist):
        return flist
    else:
        skip_count = len(flist) - i
        return drop(flist, skip_count)

def drop(lst, n):
    if not lst or n <= 0:
        return lst
    else:
        return drop(lst[1:], n - 1)

#############################################################################

\end{minted}

\paragraph{Code generated without ILP}
\begin{minted}[breaklines, fontsize=\small]{markdown}
 Python code generated by LLMs:

#####################################################################

ChatGPT 4:

def take_right(flist, i):
    """
    Returns the last i elements of flist.

    """
    # If i <= 0, return an empty list.
    if i <= 0:
        return []
    
    # If the list is empty or i >= the list length, return the full list.
    if not flist or i >= len(flist):
        return flist
    
    # Otherwise, skip the first (len(flist) - i) elements.
    skip_count = len(flist) - i
    return flist[skip_count:]
    
##################################################################################

Claude 3.5 Sonnet:

def take_right(flist, i):
    """
    Returns the last i elements of flist.
    
    Args:
        flist: input list
        i: number of elements to take from the right
        
    Returns:
        list containing the last i elements

    """
    return flist[-i:] if i > 0 else []

###################################################################################

Llama 3.1 2B Local:

def take_right(flist, i):
    # If i is less than or equal to 0, return an empty list.
    if i <= 0:
        return []
    
    # If the list is empty or i is greater than or equal to the list length, return the list.
    if not flist or i >= len(flist):
        return flist
    
    # Otherwise, slice the list from the end to get the last i elements.
    # In Scheme, this would be equivalent to (list-tail flist (- (length flist) i)).
    # In Python, we can use list slicing to achieve this.
    return flist[-i:]

####################################################################################

Gemini 2.0:

def take_right(flist, i):
  """
  Fully based on the file, generate a function take-right flist i -> list 
  take-right returns the last i elements of flist.
  """
  if i == 0:
    return []
  
  return flist[-i:]

####################################################################################

LLama 3.3 Local:

def take_right(flist, i):
    """
    Returns the last i elements of flist.
    
    Args:
    flist (list): The input list.
    i (int): The number of elements to take from the right.
    
    Returns:
    list: The last i elements of flist.
    """
    if i > len(flist):
        raise ValueError("i cannot be greater than the length of flist")
    return flist[-i:]

#############################################################################

Qwen32B:

def take_right(flist, i):
    """
    Returns the last i elements of the list flist.

    :param flist: List from which to take elements
    :param i: Number of elements to take from the end of the list
    :return: List containing the last i elements of flist
    """
    return flist[-i:]

#############################################################################

\end{minted}